\documentclass[iop,apj]{emulateapj}

\usepackage{natbib}
\defcitealias{2001ApJ...556..121K}{Ke01}
\defcitealias{2006MNRAS.372..961K}{Ke06}
\defcitealias{2003MNRAS.346.1055K}{Ka03}

\shorttitle{AGNs with a Low-metallicity NLR}
\shortauthors{Kawasaki et al.}

\begin{document}

\title{Active Galactic Nuclei with a Low-Metallicity Narrow-Line Region}

\author{
   Kota Kawasaki\altaffilmark{1}, 
   Tohru Nagao\altaffilmark{2}, 
   Yoshiki Toba\altaffilmark{3}, 
   Koki Terao\altaffilmark{1}, and 
   Kenta Matsuoka\altaffilmark{4, 5, 6}
}
\email{kawasaki@cosmos.phys.sci.ehime-u.ac.jp}
\altaffiltext{1}{
   Department of Physics, Graduate School of Science and Engineering, 
   Ehime University, 2-5 Bunkyo-cho, Matsuyama, Ehime 790-8577, Japan} 
\altaffiltext{2}{
   Research Center for Space and Cosmic Evolution, 
   Ehime University, 2-5 Bunkyo-cho, Matsuyama, Ehime 790-8577, Japan} 
\altaffiltext{3}{
   Academia Sinica Institute of Astronomy and Astrophysics,
   PO Box 23-141, Taipei 10617, Taiwan} 
\altaffiltext{4}{
   Department of Astronomy, Kyoto University, Kitashirakawa-
   Oiwake-cho, Sakyo-ku, Kyoto 606-8502, Japan}  
 \altaffiltext{5}{
   Dipartimento di Fisica e Astronomia, Universit$\grave{a}$ di Firenze, Via G.
   Sansone 1, I-50019 Sesto Fiorentino, Italy}         
 \altaffiltext{6}{
   INAF -- Osservatorio Astrofisico di Arcetri, Largo Enrico Fermi 5,
   I-50125 Firenze, Italy}

\begin{abstract}
Low-metallicity active galactic nuclei (AGNs) are interesting to study 
the early phase of the AGN evolution. 
However most AGNs are chemically matured and accordingly 
low-metallicity AGNs are extremely rare. 
One approach to search for low-metallicity AGNs systematically is utilizing 
the so-called BPT diagram that consists of the [O~{\sc iii}]$\lambda$5007/H$\beta$$\lambda4861$ and 
[N~{\sc ii}]$\lambda6584$/H$\alpha$$\lambda6563$ flux ratios. 
Specifically, photoionization models predict that low-metallicity
AGNs show a high [O~{\sc iii}]$\lambda$5007/H$\beta$$\lambda$4861 ratio
and a relatively low [N~{\sc ii}]$\lambda$6584/H$\alpha$$\lambda$6563 ratio,
that corresponds to the location between the sequence of star-forming galaxies
and that of usual AGNs on the BPT diagram (hereafter ``the BPT valley''). 
However, other populations of galaxies such as star-forming galaxies and AGNs with a high 
electron density or a high ionization parameter could be also located in the BPT
valley, not only low-metallicity AGNs.
In this paper, we examine whether most of emission-line galaxies at 
the BPT valley are low-metallicity AGNs or not. We select 70 BPT-valley objects 
from 212,866 emission line galaxies obtained by the Sloan Digital Sky Survey. 
Among the 70 BPT-valley objects, 43 objects show
firm evidence of the AGN activity; i.e., the He~{\sc ii}$\lambda$4686 emission 
and/or weak but significant broad H$\alpha$ emission.
Our analysis shows that those 43 BPT-valley
AGNs are not characterized by a very high gas density nor ionization parameter, inferring that 
at least 43 among 70 BPT-valley objects (i.e., $>60$\%) are low-metallicity AGNs. 
This suggests that the BPT diagram is an efficient tool to search for low-metallicity AGNs. 
\end{abstract}


\section{Introduction}

The active galactic nucleus (AGN) is one of the most luminous class of objects 
in the Universe, whose huge radiative energy is released through the mass 
accretion onto the supermassive black hole (SMBH). The mass of SMBHs 
($M_{\rm BH}$) is tightly correlated with the mass or the stellar velocity 
dispersion of their host galaxies \citep[e.g.,][]{1998AJ....115.2285M, 
2003ApJ...589L..21M,2013ARA&A..51..511K, 2015ApJ...813...82R}, 
implying that SMBHs and 
galaxies have evolved with closely interacting in each other (the so-called 
co-evolution of SMBHs and galaxies). However, the physics behind the 
co-evolution is still unclear. To understand the total picture of the 
co-evolution, examining the scaling relations for AGNs in the early phase of the 
co-evolution is an interesting approach since different theoretical models predict 
different redshift dependences of scaling relations \citep[e.g.,][]
{2003ApJ...583...85K, 2010MNRAS.405...29L}. One simple strategy to explore 
the early phase of the co-evolution is measuring the scaling relations at high 
redshifts, where the typical age of AGNs is much younger than low-redshift 
AGNs. Many attempts have been made for measuring the scaling relations for 
high-redshift AGNs \citep[e.g.,][]{2008A&A...478..311S, 2010ApJ...714..699W, 
2013A&A...559A..29C}, and a higher $M_{\rm BH}$ with respect to the mass 
or velocity dispersion of host galaxies has been sometimes reported 
\citep[e.g.,][]{2010ApJ...714..699W}. On the other hand, there are some reports 
claiming that such a possible evolution in the scaling relation is a result of 
observational bias through the sample selection \citep[e.g.,][]
{2011A&A...535A..87S}. Measuring the properties of AGN host galaxies at high 
redshift is generally very challenging, that prevents us from assessing the scaling
relations at high redshifts.

Another possible approach to study the early phase of the co-evolution is 
focusing on young AGNs at low redshifts, where detailed observations are much 
easier than high redshifts. In this context, low-metallicity (i.e., chemically young) 
AGNs in the low-redshift Universe are particularly interesting. However, the 
typical metallicity of AGNs inferred for broad-line regions (BLRs) and narrow-line
regions (NLRs) is high ($Z \gtrsim 2 Z_{\odot}$; e.g., 
\citealt{2006A&A...447..157N, 2009A&A...503..721M}) and low-metallicity AGNs
are very rare \citep[e.g.,][]{2008ApJ...687..133I}. \citet{2006MNRAS.371.1559G}
proposed a method to search for AGNs with a low-metallicity NLR, that utilizes 
an optical emission-line diagnostic diagram which consists of the flux ratios of
[N~{\sc ii}]$\lambda$6584/H$\alpha$$\lambda$6563 and
[O~{\sc iii}]$\lambda$5007/H$\beta$$\lambda$4861. This diagnostic diagram
was originally investigated for classifying emission-line galaxies into star-forming
galaxies and Seyfert 2 galaxies (BPT diagram, \citealt{1981PASP...93....5B}).
\citet[hereafter Ke01]{2001ApJ...556..121K} established the ``maximum'' 
starburst line in the BPT diagram by combining stellar population synthesis 
models and photoionization models. On the other hand, \citet[hereafter Ka03]
{2003MNRAS.346.1055K} derived empirical classification criteria for star-forming 
galaxies while \citet[hereafter Ke06]{2006MNRAS.372..961K} derived empirical 
classification criteria for low-ionization nuclear emission-line regions (LINERs;
\citealt{1980A&A....87..152H}), using emission-line data taken from the database 
of Sloan Digital Sky Survey (SDSS; \citealt{2000AJ....120.1579Y}).

\citet{2006MNRAS.371.1559G} pointed out that AGNs with a low-metallicity 
NLRs (i.e., characterized by the solar or sub-solar metallicity) should have a flux ratio of 
[O~{\sc iii}]$\lambda$5007/H$\beta$$\lambda$4861 as high as usual AGNs
($\sim 10^{0.5}-10^1$) but have an intermediate flux ratio of
[N~{\sc ii}]$\lambda$6584/H$\alpha$$\lambda$6563 between usual AGNs and
low-mass (i.e., low-metallicity) star-forming galaxies ($\sim 10^{-1}-10^{-0.5}$). 
This is because the nitrogen relative abundance is in proportion to the metallicity 
due to its nature as a secondary element \citep[e.g.,][]{1998ApJ...497L...1V}. In 
the BPT diagram, there are only few objects located at the region characterized 
by a high flux ratio of [O~{\sc iii}]$\lambda$5007/H$\beta$$\lambda$4861 and 
an intermediate flux ratio of [N~{\sc ii}]$\lambda$6584/H$\alpha$$\lambda$6563
(hereafter ``BPT valley''; see Figure~\ref{BPT_diagram}). 
\citet{2006MNRAS.371.1559G} specifically focused on AGNs with a low-mass
host galaxy (i.e., $M_{\rm host} < 10^{10} M_\odot$), and then they selected 
low-metallicity AGNs using another diagnostic
diagram that consists of [N~{\sc ii}]$\lambda$6584/[O~{\sc ii}]$\lambda$3727 and
[O~{\sc iii}]$\lambda$5007/[O~{\sc ii}]$\lambda$3727 flux ratios. However, it is not
clear whether low-metallicity AGNs should be always found in a sample of AGNs with a
low-mass host galaxy. Also, the method adopted by \citet{2006MNRAS.371.1559G}
requires a wide wavelength coverage ($\lambda_{\rm rest} \sim 3700-6600$ \AA), 
that is not convenient for future applications to expand the search of low-metallicity
AGNs toward the high-redshift Universe. 

Therefore, we focus on BPT-valley selection (requiring a moderately narrow 
wavelength coverage; $\lambda_{\rm rest} \sim 4800-6600$ \AA) to select 
low-metallicity AGNs without any host-mass cut.
However, there is a potentially serious problem in the BPT-valley selection for 
identifying low-metallicity AGNs. 
That is, not only low-metallicity AGNs are located in the BPT valley. 
As \citet{2013ApJ...774..100K} showed, star-forming galaxies with 
a very hard radiation field or high-density H~{\sc ii} regions are expected 
to be seen in the BPT valley (see also, e.g., \citealt[][]{2014ApJ...787..120S}). 
Also, star-forming galaxies with a high ionization parameter 
\citep[e.g.,][]{2014ApJ...795..165S, 2015PASJ...67...80H}, a high nitrogen-to-oxygen 
abundance ratio (N/O; e.g., \citealt[][]{2014ApJ...785..153M, 2015ApJ...801...88S, 2015PASJ...67..102Y, 
2016arXiv160503436K}), 
or shocks \citep[e.g.,][]{2014ApJ...781...21N} are also expected to be seen in the BPT valley. 
Not only star-forming galaxies, AGNs with a high electron density or high ionization parameter 
(i.e., not characterized by a low metallicity) could be also seen in the BPT valley 
\citep[e.g.,][]{2001ApJ...546..744N}. 
Therefore, it is not completely clear whether the BPT-valley objects are really low-metallicity AGNs 
and whether the BPT diagram is a useful tool to search for low-metallicity AGNs. 
This problem prevents us from selecting chemically-young AGNs observationally.

In this paper, we investigate the optical spectra of BPT-valley objects for 
examining whether most of emission-line galaxies at the BPT valley are really 
low-metallicity AGNs or not. Through this examination, it will be tested whether 
the optical BPT diagram is an efficient and appropriate method to search for 
low-metallicity AGNs. In Section 2, we present our selection procedure of the 
BPT-valley sample. In Section 3, we show how we identify BPT-valley AGNs to 
avoid contaminating star-forming galaxies at the BPT valley. In Section 4, we 
investigate gas properties of the selected BPT-valley AGNs such as electron 
density and ionization parameter, for examining whether the BPT-valley AGNs
are characterized by a low metallicity or not. 
In Section 5, we disccus physical properties of the BPT-valley AGNs.
Section 6 describes the summary of this work.

\section{Sample}

\begin{figure}[h]
 \centering
 \includegraphics[width=8.5cm]{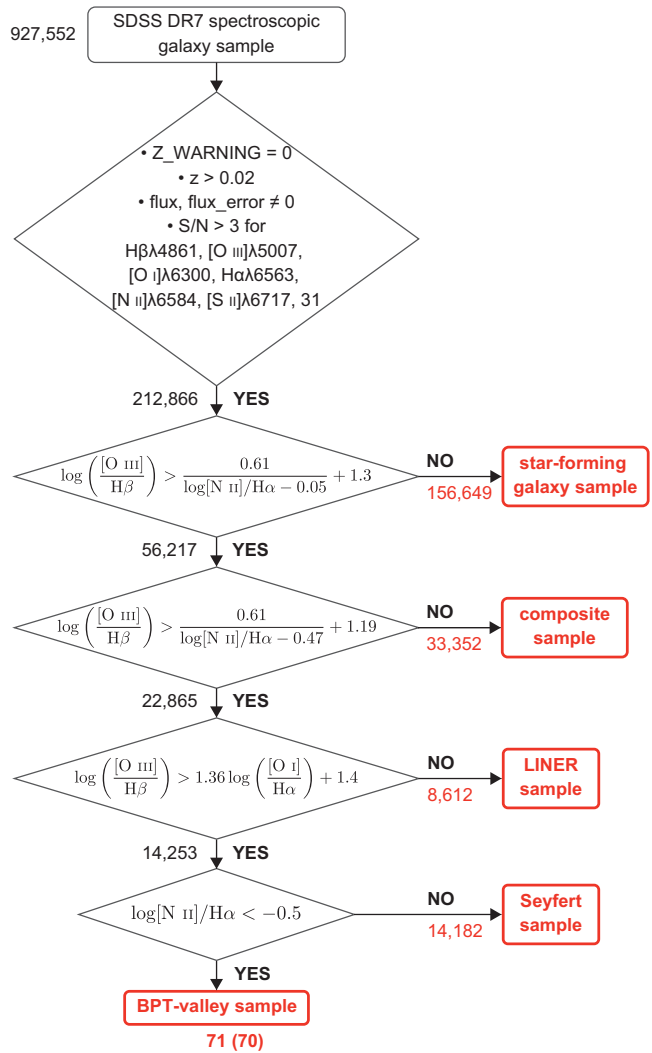}
 \caption{Flow chart of the selection of BPT-valley objects. Numbers shown in the chart denote 
 the numbers of objects at each selection stage. 
 The number shown in the parenthesis denotes the number of objects after removing the duplication.}
 \label{flow_chart} 
\end{figure}

In order to select the BPT-valley objects, we use Max-Planck-Institute for 
Astrophysics (MPA)-Johns Hopkins University (JHU) SDSS Data Release 7 
(\citealt{2009ApJS..182..543A}) galaxy catalog\footnote[1]
{http://www.mpa-garching.mpg.de/SDSS/}. 
The MPA-JHU DR7 catalog of spectral measurements contains various
spectral properties such as emission-line fluxes and their errors, based on
the analysis for 927,552 objects without showing dominant broad Balmer lines 
(i.e., star-forming galaxies, composite galaxies, LINERs, and type-2 Seyfert galaxies) 
in the SDSS DR7.  
Our sample selection is based on the following procedure (the flow chart of our 
sample selection process is shown in Figure~\ref{flow_chart}).

First, we select the initial sample according to the following criteria. 
We require the reliable redshift measurement (i.e., $z_{\rm warning} = 0$) and also $z>$ 0.02. 
This redshift limit is required to cover [O~{\sc ii}]$\lambda$3727.
This results in 906,761 galaxies.
Then we require a signal-to-noise ratio (S/N) $>$ 3 for some key emission lines, i.e., 
H$\beta$$\lambda$4861, [O~{\sc iii}]$\lambda$5007, 
[O~{\sc i}]$\lambda$6300, H$\alpha$$\lambda$6563, 
[N~{\sc ii}]$\lambda$6584 and [S~{\sc ii}]$\lambda \lambda$6717, 31 (212,866 galaxies).

\begin{figure}[h]
 \centering
 \includegraphics[width=8.5cm]{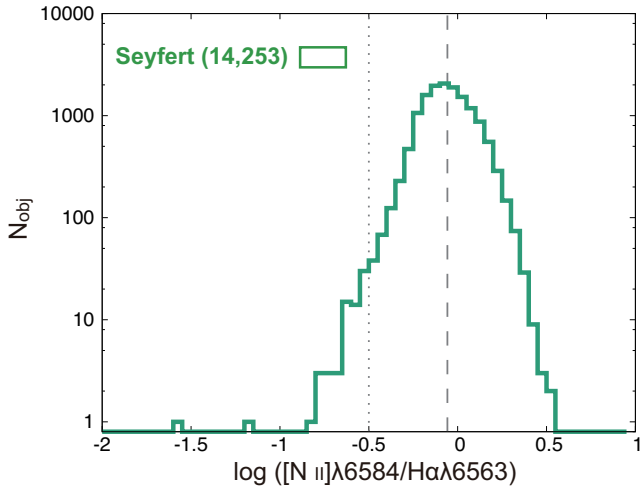}
 \caption{
 Histogram of the [N~{\sc ii}]$\lambda$6584/H$\alpha$$\lambda$6563 flux ratio of Seyfert galaxies.
 Dashed line denotes the average of log ([N~{\sc ii}]$\lambda6584$/H$\alpha$$\lambda6563$) $= -0.058$, 
 while dotted line denotes the threshold of log ([N~{\sc ii}]$\lambda6584$/H$\alpha$$\lambda6563$) $= -0.5$ 
 to select BPT-valley objects.
 }
 \label{NII_Ha} 
\end{figure}

\begin{figure}[h]
 \centering
 \includegraphics[width=8.5cm]{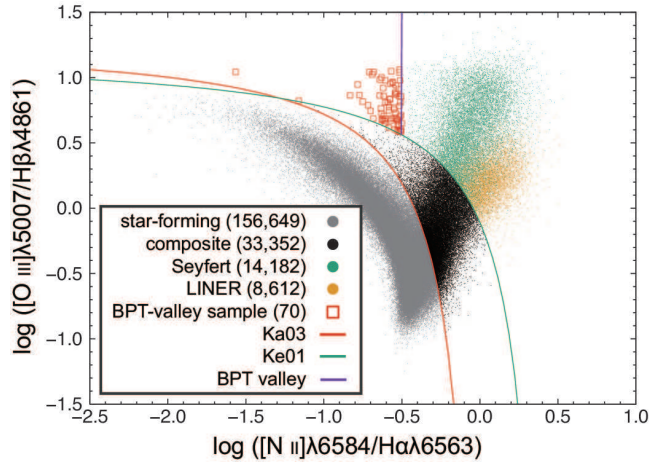}
 \caption{The BPT diagram ([N~{\sc ii}]$\lambda$6584/H$\alpha$$\lambda$6563 versus 
 [O~{\sc iii}]$\lambda$5007/H$\beta$$\lambda$4861), showing the BPT-valley sample 
 (red square) among the SDSS DR7 emission-line objects. 
 The green solid line is the Ke01 extreme starburst criterion, 
 while the red solid line denotes the criterion for separating star-forming galaxies and 
 composite galaxies (Ka03). 
 The violet solid line is the BPT valley criterion which is defined in this paper.
 The numbers of various galaxy populations are shown in the parenthesis 
 in the lower-left box.}
 \label{BPT_diagram} 
\end{figure}

Next, we classify these 212,866 galaxies and extract the BPT-valley sample 
according to the following steps. 
\begin{enumerate}
\item 
   Using the \citetalias{2003MNRAS.346.1055K} empirical line,
   \begin{eqnarray}
   \log \left( \frac{\rm [O\ {\scriptstyle III}]}{\rm H\beta} \right) 
   > \frac{0.61}{\rm log ([N\ {\scriptstyle II}]/H\alpha) -0.05}+1.3,
   \end{eqnarray}
   for removing usual star-forming galaxies (56,217 galaxies).
\item 
   Using the \citetalias{2001ApJ...556..121K} maximum starburst line,
   \begin{eqnarray}
   \log \left( \frac{\rm [O\ {\scriptstyle III}]}{\rm H\beta} \right) > 
   \frac{0.61}{\rm log ([N\ {\scriptstyle II}]/H\alpha) -0.47}+1.19,
   \end{eqnarray}
   for removing so-called composite galaxies (22,865 galaxies).
\item 
   Using the \citetalias{2006MNRAS.372..961K} empirical criterion,
   \begin{eqnarray}
   \log \left( \frac{\rm [O\ {\scriptstyle III}]}{\rm H\beta} \right) > 
   1.36\log \left( \frac{\rm [O\ {\scriptstyle I}]}{\rm H\alpha} \right) + 1.4, 
   \end{eqnarray}
   for obtaining Seyfert sample by removing LINERs (14,253 galaxies).
\item  
   Adopting the following criterion,
   \begin{eqnarray}
   \log \left( \frac{\rm [N\ {\scriptstyle II}]}{\rm H\alpha} \right) < -0.5,
   \label{BPT_valley}
   \end{eqnarray}
   for finally selecting the BPT-valley sample (71 galaxies).
\end{enumerate}
Note that 1 object in the 71 BPT-valley objects was observed twice and 
duplicated in the final sample, i.e., the final BPT-valley sample consists of 70 objects. 
The BPT-valley criterion (Equation~\ref{BPT_valley}) is determined empirically, 
by taking account of the frequency distribution of the [N~{\sc ii}]$\lambda6584$/H$\alpha$$\lambda6563$ 
flux ratio of Seyfert galaxies. 
Figure~\ref{NII_Ha} shows the [N~{\sc ii}]$\lambda6584$/H$\alpha$$\lambda6563$ frequency distribution of 
Seyfert galaxies, 
where the average and standard deviation of the logarithmic [N~{\sc ii}]$\lambda6584$/H$\alpha$$\lambda6563$ 
flux ratio are $-0.058$ and $0.145$, respectively.
Accordingly, the $3\ \sigma$ bounding from the average value is $-0.493$, 
and therefore we adopt the threshold to categorize BPT-valley objects as described 
by Equation~\ref{BPT_valley}.
Figure~\ref{BPT_diagram} shows the finally selected 70 BPT-valley objects 
in the BPT diagram that consists of
[N~{\sc ii}]$\lambda$6584/H$\alpha$$\lambda$6563 versus 
[O~{\sc iii}]$\lambda$5007/H$\beta$$\lambda$4861.
Table 1 shows the basic properties of the selected BPT-valley objects.

\begin{deluxetable*}{ccrrrrrrrrc}
\tabletypesize{\scriptsize}
\tablecaption{The BPT-valley sample}
\tablewidth{0pt}
\tablehead{
\colhead{ID} & \colhead{SDSS Name} & \colhead{Plate} & \colhead{MJD} & \colhead{Fiber} &
\colhead{$z$} & \colhead{H$\beta$$\lambda$4861} & \colhead{[O~{\sc iii}]$\lambda$5007} &
\colhead{[O~{\sc i}]$\lambda$6300} & \colhead{H$\alpha$$\lambda$6563} &
\colhead{[N~{\sc ii}]$\lambda$6584}\\
\colhead{(1)} & \colhead{(2)} & \colhead{(3)} & \colhead{(4)} & \colhead{(5)} & 
\colhead{(6)} & \colhead{(7)} & \colhead{(8)} & \colhead{(9)} & \colhead{(10)} & 
\colhead{(11)}
}
\startdata
1.......... & SDSS J102310.97$-$002810.8 & 0272 & 51941 & 0238 & 0.11274 & 253.97  & 2279.29 & 125.19 & 881.10   & 235.20  \\
2.......... & SDSS J111006.26$-$010116.5 & 0278 & 51900 & 0096 & 0.10949 & 382.92  & 1837.07 & 129.44 & 1484.61  & 438.76  \\
3.......... & SDSS J230321.73+011056.4 & 0380 & 51792 & 0565 & 0.18136 & 235.38  & 1692.71 & 12.29  & 816.35   & 247.33  \\
4.......... & SDSS J024825.26$-$002541.4 & 0409 & 51871 & 0150 & 0.02467 & 10.35   & 41.33   & 4.71   & 47.65    & 14.92   \\
5.......... & SDSS J073506.37+393300.8 & 0432 & 51884 & 0316 & 0.03479 & 29.46   & 163.57  & 8.55   & 109.51   & 21.85   \\
6.......... & SDSS J023310.78$-$074813.4 & 0455 & 51909 & 0388 & 0.03097 & 97.71   & 707.20  & 12.83  & 433.29   & 104.88  \\
7.......... & SDSS J092907.78+002637.3 & 0475 & 51965 & 0205 & 0.11732 & 259.86  & 2602.18 & 56.12  & 1173.36  & 274.70  \\
8.......... & SDSS J090613.76+561015.1 & 0483 & 51902 & 0016 & 0.04668 & 188.36  & 1144.93 & 73.96  & 621.46   & 189.46  \\
9.......... & SDSS J144328.78+044022.0 & 0587 & 52026 & 0374 & 0.11411 & 36.11   & 300.13  & 3.37   & 137.53   & 32.98   \\
10......... & SDSS J114825.71+643545.0 & 0598 & 52316 & 0189 & 0.04169 & 349.08  & 1347.75 & 60.17  & 1302.50  & 392.24  \\
11......... & SDSS J213439.57$-$071641.9 & 0641 & 52199 & 0487 & 0.06377 & 215.60  & 2079.35 & 78.83  & 800.93   & 182.94  \\
12......... & SDSS J001050.35$-$010257.4 & 0686 & 52519 & 0020 & 0.11299 & 223.89  & 1669.57 & 109.74 & 940.89   & 275.90  \\
13......... & SDSS J124738.52+621243.1 & 0781 & 52373 & 0076 & 0.12112 & 171.82  & 915.80  & 33.03  & 697.15   & 162.91  \\
14......... & SDSS J131659.37+035319.9 & 0851 & 52376 & 0219 & 0.04541 & 127.05  & 1399.96 & 80.91  & 698.40   & 136.73  \\
15......... & SDSS J115908.55+525823.1 & 0881 & 52368 & 0623 & 0.06644 & 392.77  & 2812.16 & 52.13  & 1218.45  & 379.53  \\
16......... & SDSS J104632.21+543559.7 & 0906 & 52368 & 0169 & 0.14475 & 83.71   & 938.64  & 19.52  & 375.92   & 101.46  \\
17......... & SDSS J104600.36+061632.0 & 1000 & 52643 & 0035 & 0.18447 & 146.67  & 637.74  & 31.28  & 586.86   & 177.72  \\
18......... & SDSS J101945.65+520608.6 & 1008 & 52707 & 0378 & 0.06492 & 29.60   & 115.17  & 6.75   & 92.29    & 28.00   \\
19......... & SDSS J205111.11+000913.2 & 1023 & 52818 & 0393 & 0.06644 & 7.37    & 44.11   & 4.92   & 37.11    & 9.68    \\
20......... & SDSS J214930.43+010509.4 & 1031 & 53172 & 0369 & 0.11399 & 28.14   & 210.35  & 3.43   & 109.29   & 31.13   \\
21......... & SDSS J081653.27+285423.1 & 1206 & 52670 & 0530 & 0.05740 & 246.45  & 1610.19 & 131.99 & 1145.34  & 203.60  \\
22......... & SDSS J095319.42+422912.2 & 1217 & 52672 & 0349 & 0.22343 & 207.67  & 1835.79 & 83.29  & 698.87   & 115.09  \\
23......... & SDSS J110504.94+101623.5 & 1221 & 52751 & 0359 & 0.02076 & 104.02  & 466.39  & 29.14  & 481.20   & 151.62  \\
24......... & SDSS J114440.53+102429.3 & 1226 & 52734 & 0435 & 0.12688 & 51.67   & 304.73  & 39.93  & 241.38   & 67.83   \\
25......... & SDSS J124110.10+104143.7 & 1233 & 52734 & 0611 & 0.15613 & 214.49  & 1004.84 & 29.37  & 703.65   & 188.71  \\
26......... & SDSS J092620.42+352250.3 & 1274 & 52995 & 0147 & 0.24729 & 118.98  & 868.18  & 81.24  & 513.89   & 75.74   \\
27......... & SDSS J131756.07+491531.3 & 1282 & 52759 & 0390 & 0.09231 & 215.30  & 2050.51 & 105.88 & 711.65   & 176.30  \\
28......... & SDSS J090107.41+085459.2 & 1300 & 52973 & 0335 & 0.08380 & 172.61  & 1116.69 & 153.30 & 822.23   & 210.37  \\
29......... & SDSS J120134.05+581421.1 & 1313 & 52790 & 0527 & 0.04636 & 28.83   & 136.20  & 5.78   & 87.43    & 20.49   \\
30......... & SDSS J152723.47+334919.1 & 1354 & 52814 & 0044 & 0.09116 & 44.63   & 173.46  & 24.18  & 254.04   & 79.05   \\
31......... & SDSS J112314.89+431208.7 & 1365 & 53062 & 0119 & 0.08005 & 56.52   & 226.87  & 33.20  & 183.08   & 52.19   \\
32......... & SDSS J152328.09+313655.6 & 1387 & 53118 & 0210 & 0.06850 & 382.47  & 1707.84 & 67.32  & 1300.98  & 322.89  \\
33......... & SDSS J120900.89+422830.9 & 1448 & 53120 & 0075 & 0.02364 & 100.07  & 614.24  & 24.50  & 320.70   & 61.56   \\
34......... & SDSS J121839.40+470627.6 & 1451 & 53117 & 0190 & 0.09389 & 478.35  & 5046.81 & 101.68 & 1861.68  & 380.45  \\
35......... & SDSS J005231.29$-$011525.2 & 1496 & 52883 & 0089 & 0.13485 & 251.99  & 2400.52 & 55.68  & 826.34   & 254.44  \\
36......... & SDSS J011341.11+010608.5 & 1499 & 53001 & 0522 & 0.28090 & 191.99  & 2118.06 & 38.27  & 779.72   & 162.79  \\
37......... & SDSS J001901.52+003931.8 & 1542 & 53734 & 0375 & 0.09669 & 58.06   & 296.69  & 15.99  & 242.69   & 58.75   \\
38......... & SDSS J034019.39+002530.6 & 1632 & 52996 & 0467 & 0.35296 & 40.58   & 375.40  & 10.78  & 167.65   & 46.21   \\
39......... & SDSS J032224.64+401119.8 & 1666 & 52991 & 0048 & 0.02608 & 121.08  & 1388.24 & 50.70  & 428.76   & 130.77  \\
40......... & SDSS J135855.82+493414.1 & 1670 & 53438 & 0061 & 0.11592 & 56.67   & 385.56  & 14.74  & 189.76   & 50.29   \\
41......... & SDSS J160452.78+344540.4 & 1682 & 53173 & 0201 & 0.05493 & 87.33   & 437.20  & 37.34  & 364.99   & 111.94  \\
42......... & SDSS J132011.71+125940.9 & 1698 & 53146 & 0327 & 0.11398 & 25.43   & 174.93  & 4.92   & 92.99    & 24.68   \\
43......... & SDSS J143523.42+100704.1 & 1711 & 53535 & 0306 & 0.03122 & 128.18  & 530.61  & 15.31  & 473.30   & 123.16  \\
44......... & SDSS J072637.94+394557.8 & 1733 & 53047 & 0326 & 0.11141 & 505.82  & 3357.26 & 23.23  & 1744.03  & 120.42  \\
45......... & SDSS J095914.76+125916.4 & 1744 & 53055 & 0385 & 0.03432 & 1298.59 & 8418.94 & 361.09 & 4396.86  & 1088.11 \\
46......... & SDSS J113714.22+145917.2 & 1755 & 53386 & 0463 & 0.03484 & 74.19   & 364.47  & 25.99  & 276.40   & 77.15   \\
47......... & SDSS J120847.79+135906.7 & 1764 & 53467 & 0013 & 0.29030 & 136.33  & 659.31  & 28.11  & 554.94   & 137.92  \\
48......... & SDSS J135429.05+132757.2 & 1777 & 53857 & 0076 & 0.06332 & 312.80  & 3422.56 & 111.50 & 1005.37  & 306.94  \\
49......... & SDSS J130431.99+061616.7 & 1794 & 54504 & 0046 & 0.06283 & 184.84  & 1202.46 & 41.38  & 702.29   & 217.24  \\
50......... & SDSS J134316.52+101440.1 & 1804 & 53886 & 0433 & 0.08132 & 186.67  & 960.51  & 170.90 & 635.68   & 198.21  \\
51......... & SDSS J160032.89+052608.8 & 1822 & 53172 & 0012 & 0.11653 & 281.02  & 1630.88 & 69.22  & 1204.67  & 280.09  \\
52......... & SDSS J081212.84+541539.8 & 1871 & 53384 & 0060 & 0.04417 & 93.10   & 809.73  & 34.28  & 326.93   & 94.82   \\
53......... & SDSS J084038.99+245101.6 & 1931 & 53358 & 0396 & 0.04334 & 137.37  & 770.11  & 39.75  & 579.80   & 151.50  \\
54......... & SDSS J122451.88+360535.4 & 2003 & 53442 & 0112 & 0.15094 & 25.95   & 148.22  & 11.11  & 126.62   & 35.77   \\
55......... & SDSS J134237.37+273251.3 & 2017 & 53474 & 0127 & 0.04947 & 12.67   & 97.69   & 10.27  & 60.02    & 17.32   \\
56......... & SDSS J140952.03+244334.6 & 2128 & 53800 & 0358 & 0.05215 & 45.42   & 220.76  & 12.29  & 198.30   & 41.19   \\
57......... & SDSS J142535.21+314027.1 & 2129 & 54252 & 0618 & 0.03324 & 91.61   & 362.11  & 42.31  & 323.51   & 95.49   \\
58......... & SDSS J145505.97+211121.1 & 2148 & 54526 & 0122 & 0.06751 & 82.30   & 441.80  & 12.31  & 437.94   & 126.61  \\
59......... & SDSS J083200.51+191205.8 & 2275 & 53709 & 0472 & 0.03753 & 549.64  & 6069.28 & 42.46  & 15422.57 & 419.58  \\
60......... & SDSS J103731.01+280626.9 & 2356 & 53786 & 0468 & 0.04263 & 54.57   & 447.10  & 24.55  & 216.95   & 65.48   \\
61......... & SDSS J104403.52+282628.3 & 2356 & 53786 & 0618 & 0.16286 & 225.46  & 1047.00 & 17.43  & 794.20   & 193.30  \\
62......... & SDSS J104724.40+204433.5 & 2478 & 54097 & 0541 & 0.26515 & 102.51  & 751.70  & 37.17  & 391.58   & 66.16   \\
63......... & SDSS J160635.22+142201.9 & 2524 & 54568 & 0498 & 0.03245 & 162.66  & 621.83  & 41.72  & 517.08   & 160.44  \\
64......... & SDSS J171901.28+643830.8 & 2561 & 54597 & 0345 & 0.08954 & 152.42  & 709.97  & 21.69  & 586.95   & 174.32  \\
65......... & SDSS J084658.44+111457.5 & 2574 & 54084 & 0382 & 0.06296 & 130.82  & 638.15  & 41.04  & 557.91   & 161.29  \\
66......... & SDSS J095745.49+152350.6 & 2584 & 54153 & 0442 & 0.05183 & 96.42   & 702.55  & 24.58  & 514.46   & 117.37  \\
67......... & SDSS J133014.91+242153.9 & 2665 & 54232 & 0388 & 0.07151 & 50.49   & 244.84  & 15.74  & 284.46   & 76.98   \\
68......... & SDSS J135007.07+164227.2 & 2742 & 54233 & 0551 & 0.13043 & 99.01   & 903.05  & 38.82  & 495.52   & 137.00  \\
69......... & SDSS J153941.67+171421.9 & 2795 & 54563 & 0509 & 0.04583 & 157.66  & 758.85  & 14.76  & 500.33   & 119.40  \\
70......... & SDSS J143730.46+620649.4 & 2947 & 54533 & 0227 & 0.21862 & 66.97  & 290.54  & 22.39  & 219.25   & 65.39   

\enddata
\tablecomments{Col. (1): Identification number assigned in this paper. 
Col. (2): Object name. 
Col. (3)--(5): Plate-MJD-Fiber ID in the SDSS observation for analyzed spectra. 
Col. (6): Redshift measured by the SDSS pipeline.
Col. (7)--(11): Emission-line fluxes in units of $10^{-17}$ $\rm erg\ s^{-1}\ cm^{-2}$.
}
\end{deluxetable*}

\section{Selection of secure-AGN sample}

As described in Section 1, the BPT-valley sample potentially includes star-forming galaxies with 
special gas properties, not only AGNs. 
Thus we first select objects showing secure evidence of the AGN from the BPT-valley sample. 
Specifically, we regard objects showing at least one of the following two features 
in their SDSS spectra as secure AGNs; (1) a broad H$\alpha$$\lambda$6563 emission, 
and (2) a He~{\sc ii}$\lambda 4686$ emission line. 
Details of the selection procedure of secure AGNs are given below.

\subsection{Broad H$\alpha$$\lambda$6563 emission line}

The velocity profile of recombination lines is a powerful tool to examine 
the presence of AGNs, since star-forming galaxies never show a velocity 
width wider than $\sim$1000 km~s$^{-1}$ in full-width at half maximum (FWHM). 
Generally the optical spectra of type-1 AGNs show broad permitted lines whose 
velocity width is $\gtrsim$ 2000 km~s$^{-1}$ emitted from BLRs. 
The origin of recombination lines with $\rm FWHM \sim 1000 - 2000$ km~s$^{-1}$ is 
not very clear, since such lines may arise at BLRs in so-called narrow-line 
Seyfert 1 galaxies (NLS1s; e.g., Osterbrock \& Pogge 1985) or 
at NLRs in type-2 AGNs with a relatively large velocity width 
(such as NGC 1068 and NGC 1275; see, e.g., \citealt{1984ApJ...281..525H, 2000ApJ...532L.101C}). 
However, in either case, the detection of recombination lines with 
$\rm FWHM > 1000$ km~s$^{-1}$ strongly suggests the presence of AGNs. 
Therefore we search for the broad H$\alpha$$\lambda$6563 component 
in the optical spectrum of the BPT-valley objects. Here we do not search for 
the broad component of the H$\beta$$\lambda$4861 emission, 
since it is intrinsically fainter than that of the H$\alpha$$\lambda$6563 emission 
and it is sometimes affected significantly by the Fe~{\sc ii} multiplet emission
\citep[e.g.,][]{2001AJ....122..549V}.

We use an IRAF routine {\tt specfit} \citep{1994ASPC...61..437K} to find the broad 
H$\alpha$$\lambda$6563 component.
Specifically, we fit the SDSS optical spectrum of the BPT-valley objects 
in the range of $\lambda_{\rm rest} = 6200-6800\ \rm \AA$ with and 
without the broad H$\alpha$$\lambda$6563 component, and examine 
whether the addition of the broad component improves the spectral fit significantly. 
The details of the fitting procedure are as follows. 
First, we fit the optical spectrum with a linear continuum component and 
single-Gaussian emission-line components for [O~{\sc i}]$\lambda$6300, 
[O~{\sc i}]$\lambda$6363, [N~{\sc ii}]$\lambda$6548, H$\alpha$$\lambda$6563, 
[N~{\sc ii}]$\lambda$6584, [S~{\sc ii}]$\lambda$6717, and [S~{\sc ii}]$\lambda$6731 
(hereafter ``nobroad fitting''). 
Here we assume that the velocity width of all emission lines is the same, 
and the relative separation of the emission lines is fixed to be the same as 
that of their laboratory wavelengths. 
The flux ratios of [O~{\sc i}]$\lambda$6300 to [O~{\sc i}]$\lambda$6363 and 
[N~{\sc ii}]$\lambda$6584 to [N~{\sc ii}]$\lambda$6548 are fixed to be 3.00 and 2.96 
respectively \citep{1983IAUS..103..143M}, and the flux ratios among the remaining 
emission lines are kept to be free. 
Then, we add a broad component for the H$\alpha$$\lambda$6563 emission to 
the nobroad fit, where the flux, wavelength center and width of this additional 
component are kept to be free (hereafter ``broad fitting''). 
Here we recognize that the additional broad component significantly improves 
the fit by the following criterion:
\begin{eqnarray}
\frac{\tilde{\chi}^2_{\rm nobroad}-\tilde{\chi}^2_{\rm broad}}{\tilde{\chi}^2_{\rm nobroad}}>0.4,
\label{chi}
\end{eqnarray}
where $\tilde{\chi}^2_{\rm nobroad}$ and $\tilde{\chi}^2_{\rm broad}$ are the reduced 
chi-square of the nobroad fitting and broad fitting, respectively. 
Note that the threshold, 0.4, is determined empirically, so that the result 
becomes consistent with the visual inspection. 
As a result, 13 BPT-valley objects with a broad component are identified from the 70 BPT-valley objects.
Figures~\ref{broad_HeII_1} and~\ref{broad_noHeII} show the SDSS spectrum with the best-fit result 
for the BPT-valley objects with a broad H$\alpha$$\lambda$6563 component. 
Figure~\ref{ID_48} shows an example of objects (ID = 48) whose fitting result does not satisfy the 
criterion defined by Equation~\ref{chi} (for this case, the improvement of the fit is slightly less than 
the threshold, 0.32).
Note that we regard object ID = 8 as an object with a broad H$\alpha$ component, 
though the FWHM of the broad H$\alpha$ component is less than 
$1000\ {\rm km\ s^{-1}}$ (Figure~\ref{ID_8}). 
This is because this object shows [Fe~{\sc vii}]$\lambda$6087 and [Fe~{\sc x}]$\lambda$6374 
lines, that are seen only when the AGN presents. Note that such high-ionization forbidden 
emission lines are preferentially seen in type-1 AGNs 
\citep[e.g.,][]{1998ApJ...497L...9M, 2000AJ....119.2605N}. 
Note that the [Fe~{\sc vii}]$\lambda$6087 line is seen in 8 objects 
while [Fe~{\sc x}]$\lambda$6374 line is seen in 3 objects 
(including ID = 8, note that 2 objects in addition to ID = 8 show both [Fe~{\sc vii}]$\lambda$6087 
and [Fe~{\sc x}]$\lambda$6374). 
The spectral properties of the BPT-valley objects with a broad H$\alpha$$\lambda$6583 
component are summarized in Table 2.
Only 1 BPT-valley object (ID = 25) shows the broad H$\beta$ component among  
the 13 BPT-valley objects showing a broad H$\alpha$ component (see Figure~\ref{broad_HeII_1}).

\subsection{He~{\sc ii}$\lambda$4686 emission line}
The presence of a He~{\sc ii}$\lambda4686$ emission line indicates the existence of 
the hard ionizing radiation since the ionization potential for 
$\rm He^{+}$ is 54.4 eV. 
This hard radiation is naturally produced by AGNs. 
Therefore, the He~{\sc ii}$\lambda4686$ emission line is a good indicator of AGNs. 
We examine whether the SDSS optical spectrum of the BPT-valley objects show 
the He~{\sc ii}$\lambda$4686 line by the visual inspection, 
since the He~{\sc ii}$\lambda$4686 information is not given in the MPA-JHU database. 
As a result, 38 BPT-valley objects with the He~{\sc ii} emission line are identified from 
the 70 BPT-valley objects. 
Some of the SDSS spectra of BPT-valley objects with the He~{\sc ii} detection are shown 
in Figures~\ref{broad_HeII_1}, while those without the He~{\sc ii} detection are shown 
in Figure~\ref{broad_noHeII}.

\begin{figure*}[t]
 \centering
 \includegraphics[width=17cm]{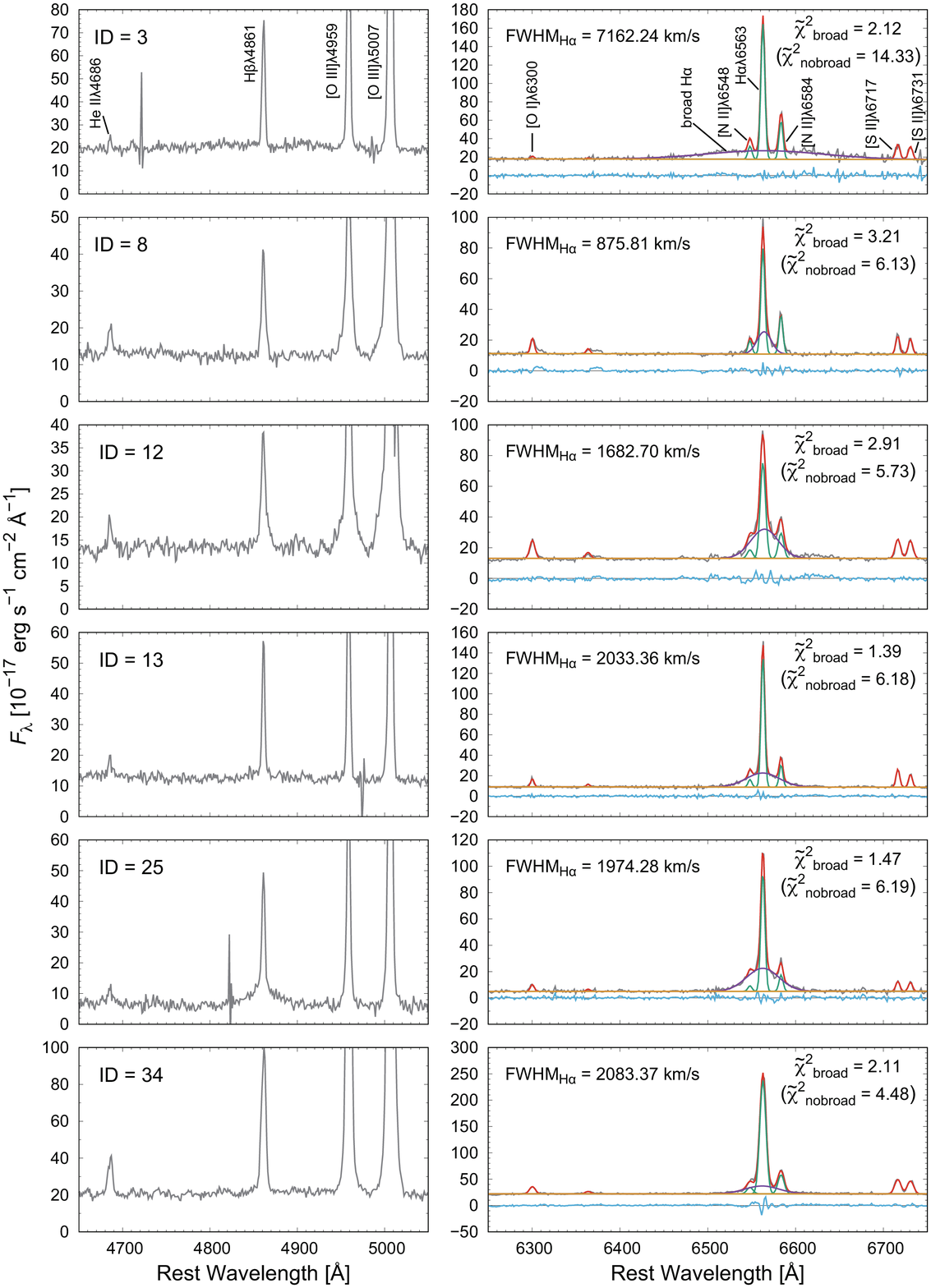}
 \vspace{2mm}
 \caption{Spectra of the BPT-valley objects showing a broad H$\alpha$ component and He~{\sc ii} emission. 
 Best-fit models are plotted in red, while the narrow-H$\alpha$+[N~{\sc ii}] Gaussian components, 
 broad H$\alpha$ component, and continuum are plotted in green, violet, and orange, respectively. 
 Residuals are plotted in blue. Reduced chi-square values are given at the upper-right side in the 
 right panels (the value before adding the broad $\alpha$ component is given in the parenthesis).
 }
 \label{broad_HeII_1} 
 \end{figure*}

\setcounter{figure}{3}
\begin{figure*}[t]
 \centering
  \includegraphics[width=17cm]{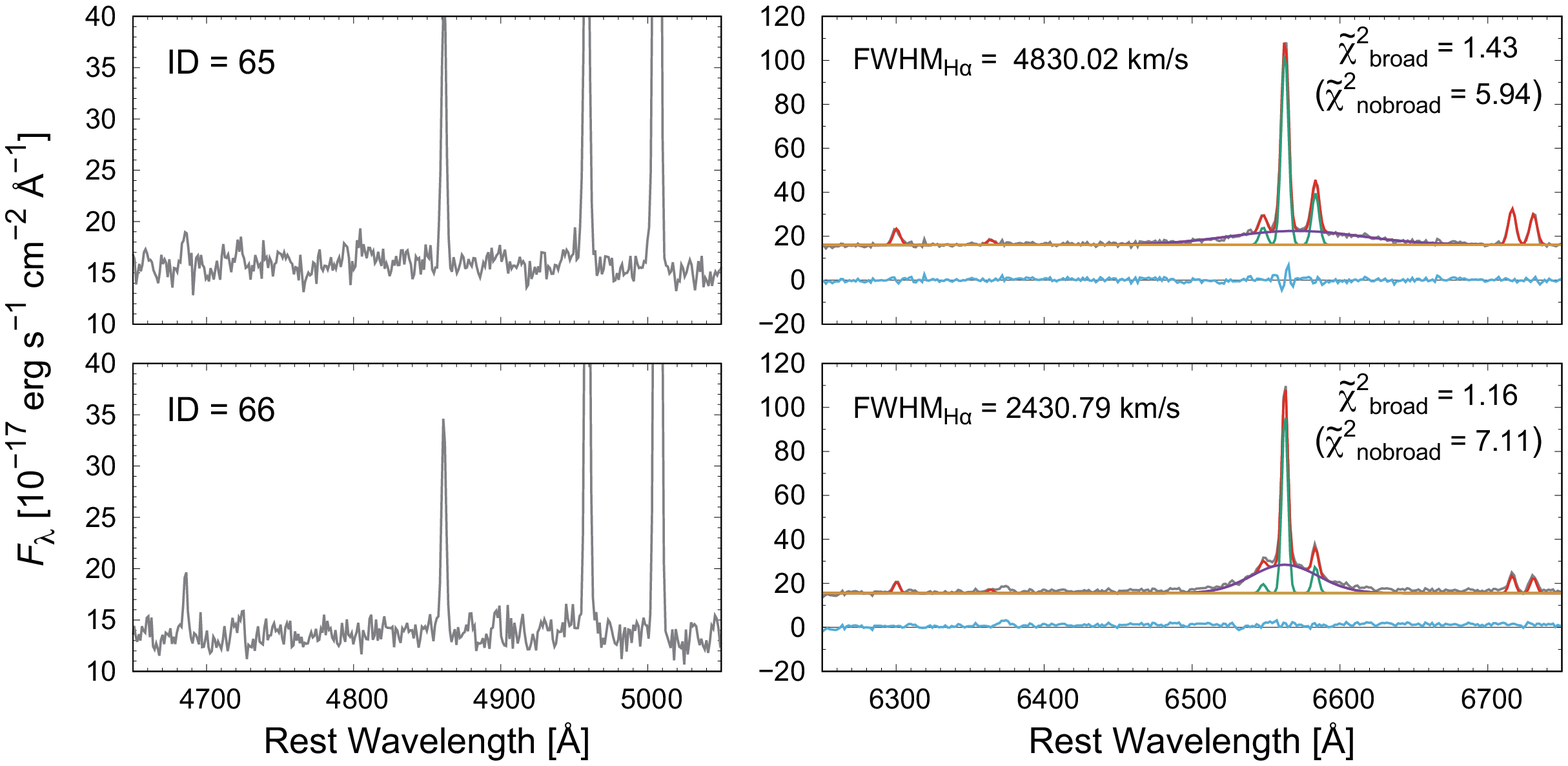}
\caption{(Continued)}
 \label{broad_HeII_2} 
\end{figure*}

\begin{figure*}[t]
 \centering
 \includegraphics[width=17.cm]{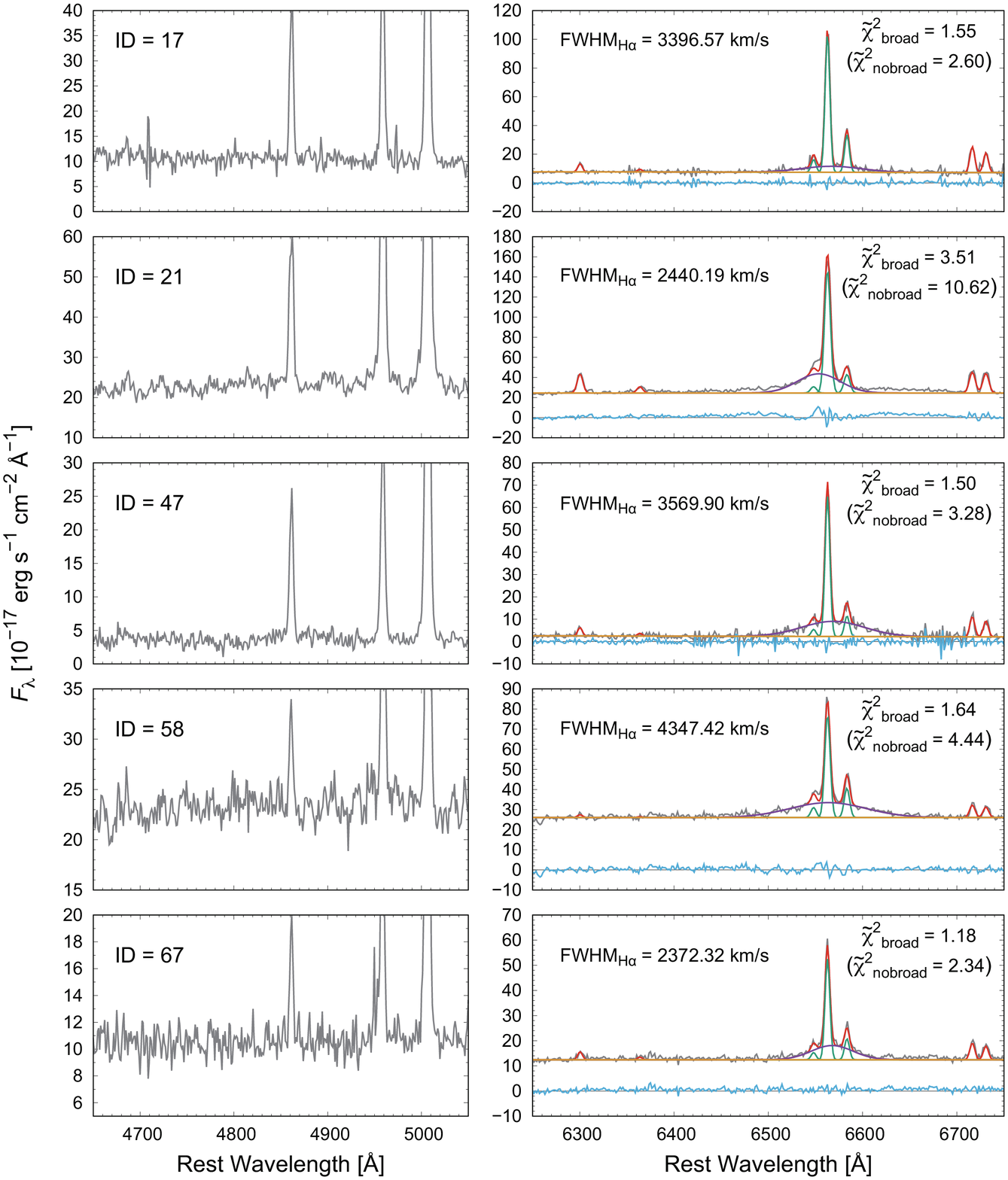}
 \vspace{2mm}
 \caption{Same as Figure~\ref{broad_HeII_1} but for objects showing the broad H$\alpha$ emission 
 but without the He~{\sc ii} line.}
 \label{broad_noHeII} 
 \end{figure*}

\begin{figure*}[t]
 \centering
 \includegraphics[width=17.cm]{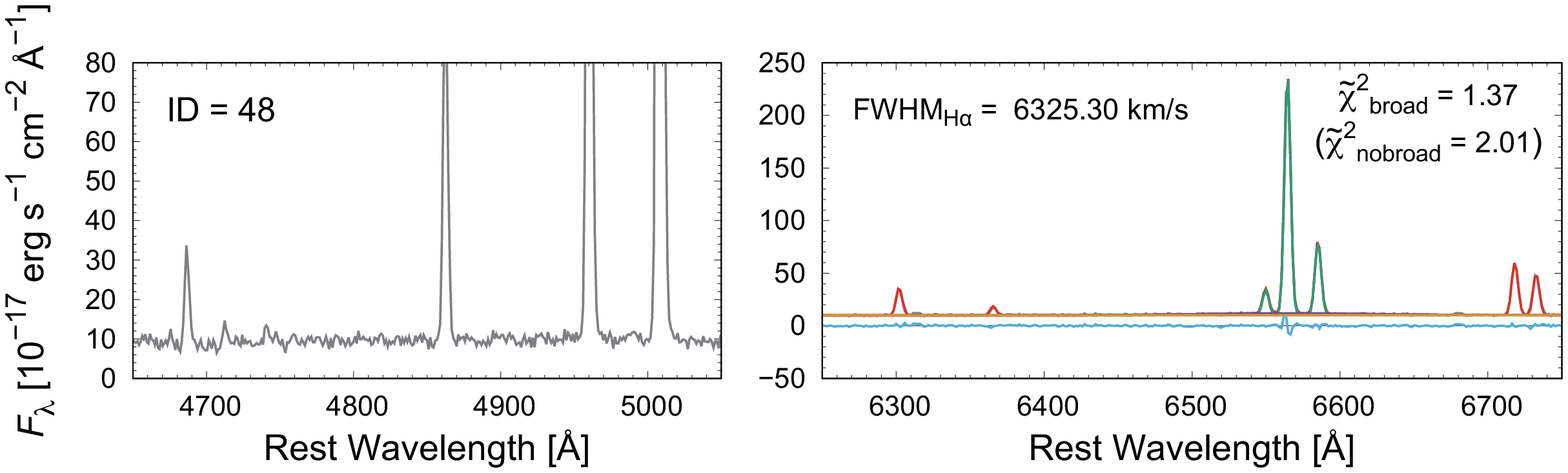}
 \vspace{2mm}
 \caption{
 Same as Figure~\ref{broad_HeII_1} but for an example of objects whose fitting result 
 does not satisfy Equation~\ref{chi}.
 }
 \label{ID_48} 
 \end{figure*}

\begin{figure}[h]
 \centering
 \hspace{0cm}
 \vspace{0.cm}
 \includegraphics[width=8.5cm]{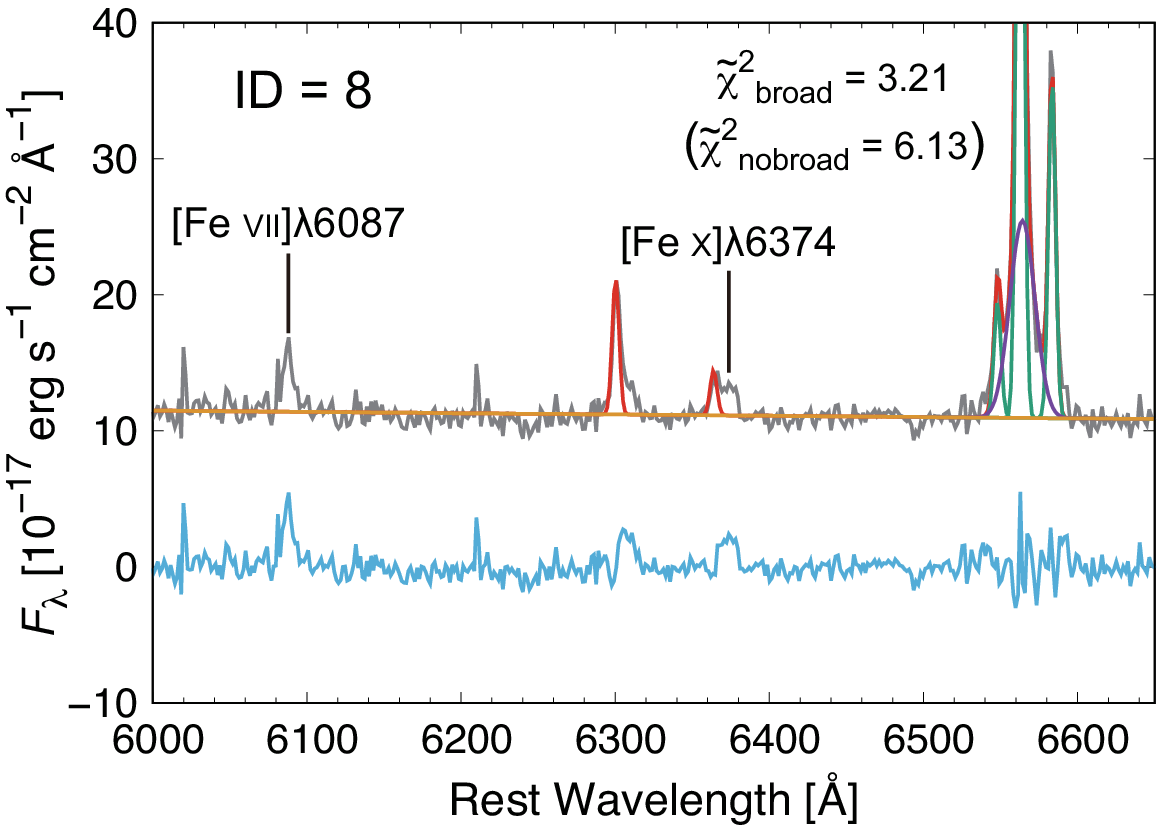}
 \vspace{0cm}
 \caption{Same as Figure~\ref{broad_HeII_1}, but for objects ID = 8, whose FWHM of broad 
 H$\alpha$ component is less than $1500\ {\rm km\ s^{-1}}$ (see Table 2). 
 High-ionization forbidden emission lines of [Fe~{\sc vii}]$\lambda$6087 and 
 [Fe~{\sc x}]$\lambda$6374 are clearly detected.
 Note that fitting range is from 6200 to 6800 $\rm \AA$ in the restframe.
 Continuum fitting is extrapolated below 6200 $\rm \AA$.}
 \label{ID_8} 
\end{figure}

\subsection{Classification result of the BPT-valley sample}

 The results of the classification of the BPT-valley objects are summarized in Table 3. 
 Among the 70 BPT-valley objects, 8 objects show both broad H$\alpha$ component and 
 He~{\sc ii} emission line, that are now confirmed to be AGNs. 
 There are 5 objects showing the broad H$\alpha$ component but without He~{\sc ii} emission line, 
 that are also regarded as AGNs. 
 The non-detection of the He~{\sc ii} line is likely due to insufficient signal-to-noise ratio, 
 since the He~{\sc ii} line is very weak. In addition, 30 objects show the He~{\sc ii} line 
 but without broad H$\alpha$ component, that are thought to be typical type-2 AGNs. 
 Here we should mention that the stellar absorption lines (mainly H$\alpha$) are not 
 considered in our fitting procedure. 
 Though the stellar H$\alpha$ absorption line could impact the narrow component of the H$\alpha$ 
 emission, the absorption effect is negligible for examining the presence of the broad H$\alpha$ 
 component. 
 This is because the equivalent width of the detected broad H$\alpha$ component is higher than 
 $20\ {\rm \AA}$ (the median value of $\rm EW_{rest}(H\alpha)_b$ is $44.74\ {\rm \AA}$, Table 2) 
 while the typical equivalent width of the stellar H$\alpha$ absorption is $\sim 2-3\ {\rm \AA}$ 
 in nearby galaxies \citep[e.g.,][]{1997ApJS..112..315H}.
 Note that the detected He~{\sc ii} line is not caused by Wolf-Rayet stars, 
 because the typical velocity width of the detected He~{\sc ii} line is not 
 broad ($\lesssim 1000\ \rm km\ s^{-1}$). 
 Therefore, at least 43 among the BPT-valley objects are regarded as AGNs. 
 There may be some additional AGNs in the remaining 27 objects, possibly 
 owing to insufficient S/N to detect any AGN indicators in their spectra. 
 Instead, some of those 27 objects could be non-AGNs, i.e., star-forming 
 galaxies with a relatively high N/O ratio or fast shocks. 
 We do not discuss further about those 27 objects since the main interests 
 of this work are on the BPT-valley AGN sample. 
 Accordingly, we conclude that at least 43 objects of the BPT-valley sample 
 (or $\sim$ 60\%, but probably more) are confirmed to be AGNs.
  
 As described in Section 3.1, at least one of the [Fe~{\sc vii}]$\lambda$6087 and 
 [Fe~{\sc x}]$\lambda$6374 lines are seen in 9 BPT-valley objects. 
 Interestingly, a large fraction of objects showing both the broad H$\alpha$ component and He~{\sc ii} 
 emission show such high-ionization iron lines (5 among 8 objects). 
 On the other hand, objects showing neither the broad H$\alpha$ component nor He~{\sc ii} emission 
 never shows those high-ionization iron lines. 
 Then, a few objects in the remaining two classes show high-ionization iron lines (4 among 35 objects). 
 This may infer that our classification is well tracing the presence of the AGN, 
 but the absence of high-ionization iron lines could be simply due to a low S/N ratio of the spectra. 
 
\begin{figure}[h]
\centering
\includegraphics[width=8.5cm]{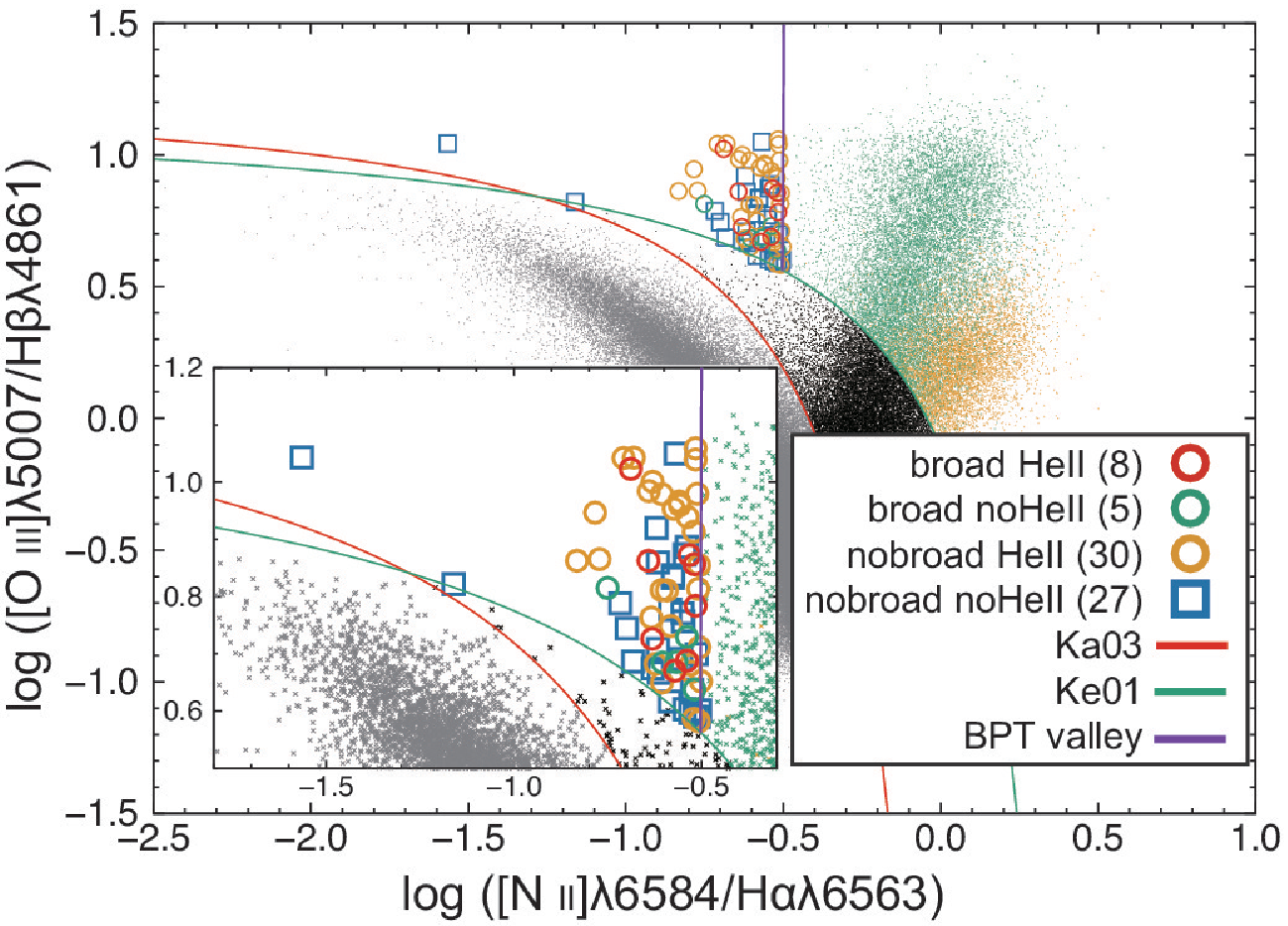}
\caption{The BPT diagram ([N~{\sc ii}]$\lambda$6584/H$\alpha$$\lambda$6563 versus 
 [O~{\sc iii}]$\lambda$5007/H$\beta$$\lambda$4861) with the classification result of the 
 BPT valley sample among the SDSS DR7 emission-line objects. 
 The numbers of various galaxy populations are shown in the parenthesis in the lower-right box.
}
\label{BPT_classification} 
\end{figure}

 Figure~\ref{BPT_classification} shows how various populations of galaxies classified in this work 
 are populated in the BPT diagram. 
 There are no significant segregation except for two BPT-valley objects whose 
 [N~{\sc ii}]$\lambda6584$/H$\alpha$$\lambda6563$ flux ratio is very low, $< 0.1$. 
 Both of these two galaxies show no broad H$\alpha$ component nor He~{\sc ii} line, 
 which is consistent with the idea that these two objects are not low-metallicity AGNs 
 but somewhat extreme low-metallicity galaxies, characterized probably by a very 
 high ionization parameter and/or very hard ionization radiation.

\begin{deluxetable*}{lrrrrr}
\tablewidth{0pt}
\tablecolumns{6} 
\tablecaption{Broad-line AGNs in the BPT valley}
\tablehead{
\colhead{ID} & \colhead{$f(\rm H\alpha)_n$} & \colhead{$f(\rm H\alpha)_b$} & \colhead{$\rm FWHM_{H\alpha}$} & \colhead{$\rm FWHM_{[S\ {\scriptscriptstyle II}]}$} & \colhead{$\rm EW_{\rm rest}(H\alpha)_b$}\\
\colhead{(1)} & \colhead{(2)} & \colhead{(3)} & \colhead{(4)} & \colhead{(5)} & \colhead{(6)}
}
\startdata
\multicolumn{6}{c}{broad H$\alpha$ and He~{\sc ii}}\\
\tableline
3......... & 991.96 & 1834.96 & 7162.24 & 245.05 & 88.10\\
8......... & 414.20 & 309.70 & 875.81\tablenotemark{1} & 247.31 & 27.02\\
12........ & 524.87 & 830.66 & 1682.70 & 324.79 & 57.25\\
13........ & 748.78 & 728.55 & 2033.36 & 223.68 & 72.83\\
25......... & 607.56 & 930.70 & 1974.28 & 248.66 & 159.88\\
34........ & 2072.41 & 797.10 & 2083.37 & 377.39 & 33.06\\
65........ & 573.27 & 761.34 & 4830.02 & 263.52 & 44.61\\
66........ & 448.38 & 768.76 & 2430.79 & 221.86 & 47.03\\
\tableline
\multicolumn{6}{c}{broad H$\alpha$ and noHe~{\sc ii}}\\
\tableline
\dataset
17........ & 719.07 & 392.29 & 3396.57  & 270.94 & 44.74\\
21........ & 1014.89 & 1145.11 & 2440.19 & 337.97 & 44.13\\
47........ & 515.66 & 726.01 & 3569.90 & 273.99 & 246.77\\
58........ & 390.35 & 801.51 & 4347.42 & 307.51 & 28.83\\
67........ & 276.51 & 335.10 & 2372.32 & 276.41 & 25.14    
\enddata
\tablenotetext{1}{Classified as an object with a broad H$\alpha$ component through the FWHM 
of the additional H$\alpha$ component is less than 1000 km s$^{-1}$ (see the main text).}
\tablecomments{
Col. (1): Identification number assigned in this paper. 
Col. (2): Flux of the nallow component of H$\alpha$ in units of $10^{-17}$ $\rm erg\ s^{-1}\ cm^{-2}$.
Col. (3): Flux of the broad component of H$\alpha$ in units of $10^{-17}$ $\rm erg\ s^{-1}\ cm^{-2}$.
Col. (4): FWHM of the broad component of H$\alpha$ in units of km $\rm s^{-1}$.
Col. (5): FWHM of the [S~{\sc ii}]$\lambda$6717 (i.e., narrow component) in units of km $\rm s^{-1}$.
Col. (6): Rest-frame equivalent width of the broad component of H$\alpha$ in units of $\rm \AA$.
}

\end{deluxetable*}

\begin{deluxetable}{lcc}
\tablecolumns{3} 
\tabletypesize{\scriptsize}
\tablewidth{0pc}
\tablecaption{Classification result of the BPT-valley sample}
\tablehead{ 
\colhead{} & \colhead{broad} & \colhead{nobroad} 
}
\startdata
    He II  & 8 & 30 \\
    noHe II   & 5 & 27 
\enddata
\end{deluxetable}

\section{Selection of Low-Metallicity AGNs}

The 43 BPT-valley objects confirmed to be AGNs are not necessarily low-metallicity AGNs, 
because AGNs with a very high electron density or very high ionization parameter are 
also expected to be populated in the BPT valley as mentioned in Section 1. 
More specifically, the [N~{\sc ii}]$\lambda$6584 emission in AGNs with a density higher than 
the critical density of the [N~{\sc ii}]$\lambda$6584 transition ($\sim$8.7 $\times 10^4$ cm$^{-3}$) 
is significantly suppressed due to the collisional de-excitation effect. 
On the other hand, a very high ionization parameter results in a higher relative ionic abundance of 
N$^{2+}$ (i.e., a lower relative ionic abundance of N$^{+}$), that results in a weaker 
[N~{\sc ii}]$\lambda$6584 emission.
Therefore, in this section, we examine whether the 43 BPT-valley AGNs are characterized 
by a very high electron density or very high ionization parameter or not, and test whether 
the AGNs in the BPT-valley are characterized by low-metallicity gas or not.

\subsection{Electron density}
The emission-line flux ratios of [S~{\sc ii}]$\lambda$6717/$\lambda$6731 and 
[O~{\sc ii}]$\lambda$3729/$\lambda$3726 are famous good indicators of the electron density 
\citep[e.g.,][]{1989agna.book.....O}.
In this work, we use the [S~{\sc ii}]$\lambda$6717/$\lambda$6731 line ratio to estimate electron density, 
because the wavelength separation of the [O~{\sc ii}] doublet is too small to be well resolved with 
the SDSS spectral resolution.
We use an IRAF routine {\tt temden} for deriving the electron density from the 
[S~{\sc ii}]$\lambda$6717/$\lambda$6731 ratio, by assuming the electron temperature of 10,000 K. 
Here we derive the electron density whose [S~{\sc ii}]$\lambda$6717/$\lambda$6731 ratio is 
within the range of 0.5--1.4. 
No BPT-valley objects show the [S~{\sc ii}]$\lambda$6717/$\lambda$6731 ratio lower than 0.5 
(i.e., the high-density limit) while 11 among the 70 BPT-valley objects show the flux ratio 
higher than 1.4 (i.e., the low-density limit). 
Among 14,252 Seyfert sample, only 12 objects show the [S~{\sc ii}]$\lambda$6716/$\lambda$6731 
ratio lower than 0.5 while 2,880 objects show the flux ratio higher than 1.4.

\begin{figure}[htbp]
 \centering
 \includegraphics[width=8.5cm]{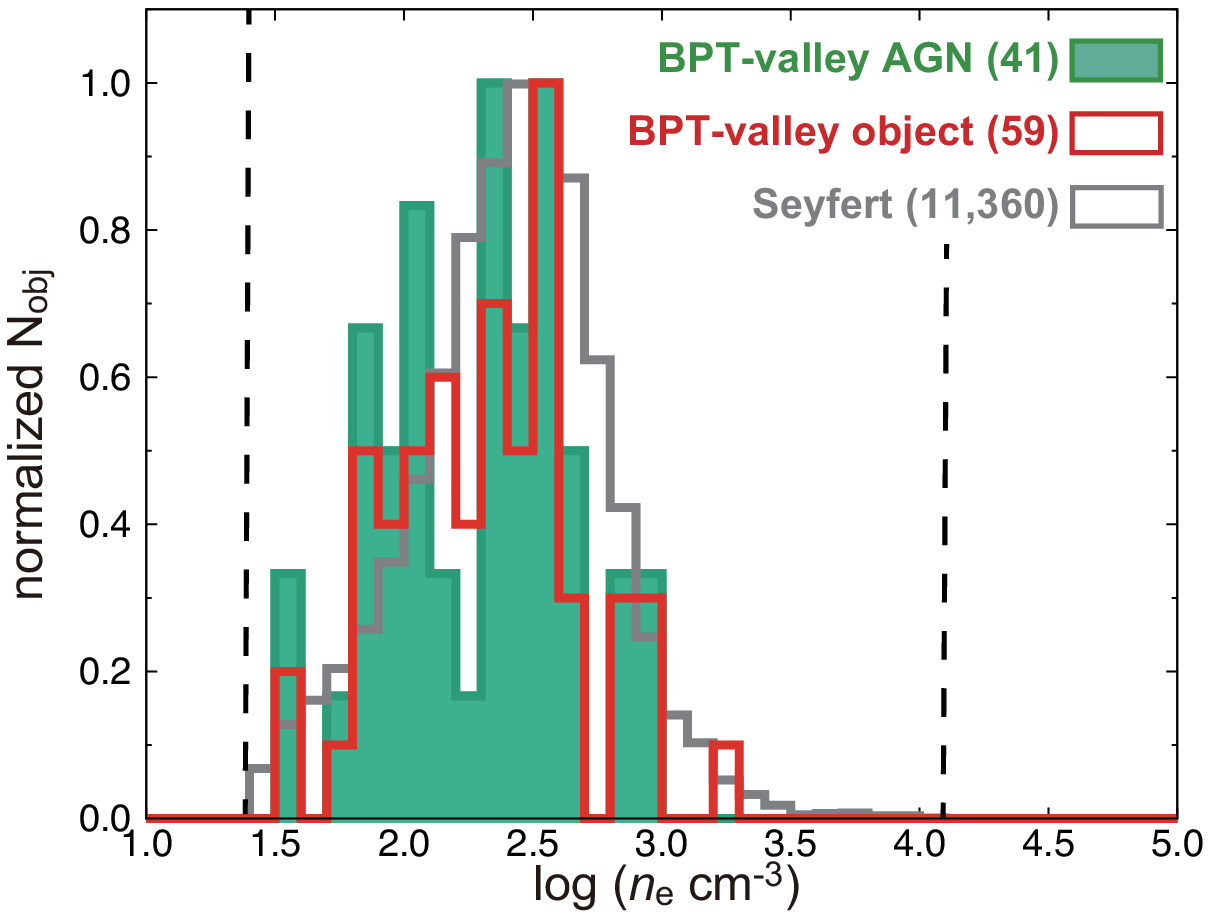}
 \caption{Histograms of the electron density of the BPT-valley AGN (filled green), 
 BPT-valley objects (open red), and Seyfert galaxies (open gray), normalized by their peak count.
 Dashed lines denote the range of the electron density measurable through the [S~{\sc ii}] doublet ratio.} 
 \label{SII_electron} 
\end{figure}

Figure~\ref{SII_electron} shows the frequency distribution of the inferred gas density for objects whose 
[S~{\sc ii}]$\lambda$6717/$\lambda$6731 ratio is within the range of 0.5--1.4; i.e.,  41 BPT-valley AGNs 
(showing a broad H$\alpha$ component and/or He~{\sc ii} emission), 59 BPT-valley objects 
(including objects without any AGN signatures), and 11,360 Seyfert galaxies. 
Here we show the histograms for both BPT-valley AGNs and BPT-valley objects, because some of 
BPT-valley objects without any AGN signatures could be also AGNs (see Section 3.3). 
The median density of the BPT-valley AGNs, BPT-valley objects, and Seyfert galaxies are 
210 cm$^{-3}$, 210 cm$^{-3}$, and 270 cm$^{-3}$, respectively. 
In order to investigate whether the frequency distribution of the gas density is statistically different 
among the samples, we apply the Kolmogorov-Smirnov (K-S) statistical test with a null hypothesis that 
the frequency distribution of the gas density of two classes of objects comes from 
the same underlying population. 
The derived K-S probability for the BPT-valley AGNs and Seyferts is 0.207, while that for 
the BPT-valley objects and Seyferts is 0.146. 
These results strongly suggest that the BPT-valley sample is not characterized by the higher gas density 
with respect to the Seyfert sample.

\subsection{Ionization parameter}
The ionization palameter is the ratio of the number density of hydrogen-ionizing photons 
to that of Hydrogen atoms. 
In order to investigate the ionization parameter, the [O~{\sc iii}]$\lambda$5007/[O~{\sc ii}]$\lambda$3727 
flux ratio is a useful indicator because this ratio does not suffer significantly from chemical properties of 
the gas in both AGNs and star-forming galaxies 
\citep[see, e.g.,][]{1997A&A...323...31K, 2002ApJ...567...73N, 2014MNRAS.442..900N}.
Note that this flux ratio is sensitive also to the gas density if the density is higher than 
the critical density of [O~{\sc ii}] ($\sim$$10^{3.5}$ cm$^{-3}$), 
but the typical density of NLRs inferred from the [S~{\sc ii}] doublet ratio is much lower than 
that as described in Section 4.1.
Though the dust reddening is not corrected to study the BPT diagram due to the small wavelength 
separation of emission-line pairs used for the BPT diagram (Section 2), we should correct for the reddening 
effect to investigate the [O~{\sc iii}]$\lambda5007$/[O~{\sc ii}]$\lambda3727$ flux ratio. 
For this correction, we assume $R_V = A_V/E(B-V) = 3.1$ and the intrinsic flux ratio 
of H$\alpha$$\lambda6584$/H$\beta$$\lambda4861$ = 3.1, and adopt the reddening curve 
of \cite{1989ApJ...345..245C}.

\begin{figure}[h]
 \centering
 \includegraphics[width=8.5cm]{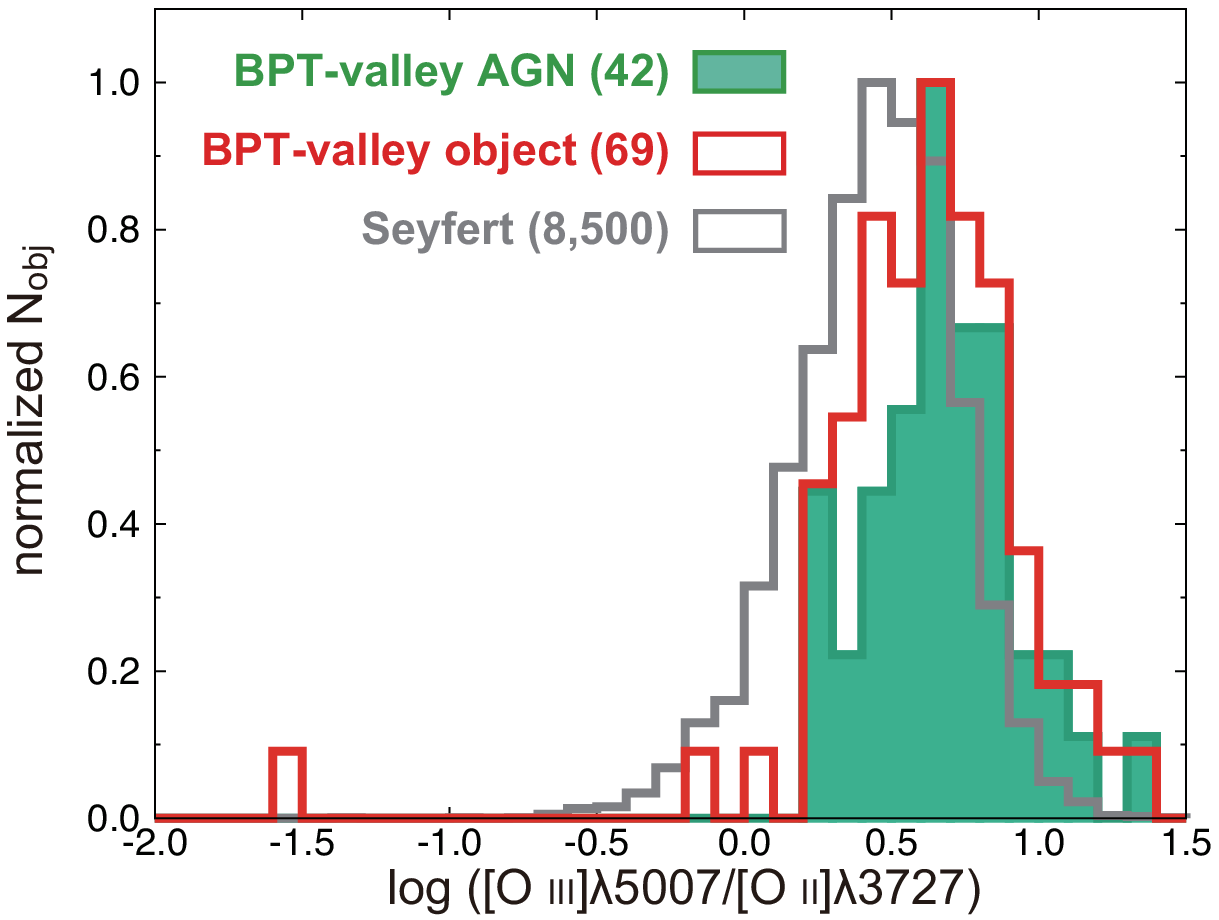}
 \caption{Same as Figure~\ref{SII_electron} but for the [O~{\sc iii}]$\lambda$5007/[O~{\sc ii}]$\lambda$3727 
 flux ratio.}
 \label{OIII_OII} 
\end{figure}

Figure~\ref{OIII_OII} shows the histogram of the [O~{\sc iii}]$\lambda$5007/
[O~{\sc ii}]$\lambda$3727 line 
ratio of the BPT-valley AGNs, BPT-valley objects, and Seyferts, with S/N([O~{\sc ii}]$\lambda$3727) $>$ 3.
Here it should be noted that the BPT-valley objects show 
log([O~{\sc iii}]$\lambda5007$/H$\beta$$\lambda4861$) $>$ 0.5 by definition, 
while the Seyfert galaxies could have much lower [O~{\sc iii}]$\lambda5007$/H$\beta$$\lambda4861$ flux ratios 
down to $\sim -0.2$. 
This may introduce a selection effect in the sense that strong [O~{\sc iii}] emitters could be selectively 
included in the BPT-valley sample. 
Therefore, for reducing this selection effect, only objects with 
log([O~{\sc iii}]$\lambda5007$/H$\beta$$\lambda4861$) $>$ 0.5 are examined for assessing 
the ionization parameter. 
After adopting this additional criterion, the numbers of the BPT-valley AGNs, BPT-valley objects, and 
Seyferts examined in Figure~\ref{OIII_OII} are 42, 69, and 8,500, respectively. 
This figure shows that the BPT-valley samples seem to show systematically higher 
[O~{\sc iii}]$\lambda5007$/[O~{\sc ii}]$\lambda3727$ flux ratios than the Seyfert sample. 
The median values of the logarithmic [O~{\sc iii}]$\lambda5007$/[O~{\sc ii}]$\lambda3727$ flux ratios of 
the BPT-valley AGNs, 
BPT-valley objects, and Seyferts are 0.67, 0.65, and 0.46, respectively. 
In order to investigate whether or not the distributions of the 
[O~{\sc iii}]$\lambda$5007/[O~{\sc ii}]$\lambda$3727 line ratio are 
statistically different between BPT-valley sample and Seyfert sample, we apply the K-S statistical test. 
The K-S probability that the underlying distribution of these two
distributions is the same is $3.925\times 10^{-6}$ for the BPT-valley AGNs and Seyferts, 
while $1.803 \times 10^{-5}$ for the BPT-valley objects and Seyferts. 
These results suggest that the BPT-valley samples have statistically higher 
[O~{\sc iii}]$\lambda5007$/[O~{\sc ii}]$\lambda3727$ flux ratios, i.e., the ionization parameter, 
than the Seyfert sample. 
Note that it is well known that low-metallicity galaxies are generally characterized by a relatively 
high ionization parameter, at least for star-forming galaxies \citep[e.g.,][]{2006A&A...459...85N}. 
It may be interesting that the BPT-valley objects show a clear edge at the lower side of the 
[O~{\sc iii}]$\lambda5007$/[O~{\sc ii}]$\lambda3727$ distribution in Figure~\ref{OIII_OII}.
However, probably this feature is not statistically significant, because the number of BPT-valley objects 
is not enough to discuss the tail of the frequency distribution of the 
[O~{\sc iii}]$\lambda5007$/[O~{\sc ii}]$\lambda3727$ flux ratio.

In the next subsection, we will examine whether or not this difference in the ionization parameter can be 
responsible for the lower [N~{\sc ii}]$\lambda6584$/H$\alpha$$\lambda6563$ ratio observed in the 
BPT-valley samples with respect to the Seyfert sample.

\subsection{Model calculations}
As shown in Section 4.2, the ionization parameter of the BPT valley sample is higher than that of 
the Seyfert sample. 
Since it is interesting to examine either the BPT-valley AGNs are characterized by a low metallicity or 
a high ionization parameter, we perform photoionization model calculations.

We perform photoionization model calculations for simulating the NLR of AGNs, 
using the code CLOUDY version 13.03 \citep{1998PASP..110..761F}. 
Here the main parameters for CLOUDY calculations are as follows 
\citep[see][for more details]{2001ApJ...546..744N}:
\begin{enumerate}
 \item The hydrogen density of the cloud ($n_{\rm H}$).
 \item The ionization parameter ($U$).
 \item The chemical composition of the gas.
 \item The shape of the input SED.
\end{enumerate}
We calculate photoionization models covering the following ranges of parameters:
$10^1\ {\rm cm^{-3}} \leq n_{\rm H} \leq 10^6\ {\rm cm^{-3}}$ and
$10^{-4} \leq U \leq 10^{-1}$. 
We set the gas-phase elemental abundance ratios to be the solar ones. 
The adopted solar abundances relative to hydrogen are taken from \cite{1989AIPC..183....1G} 
with extensions by \citet{1993oee..conf...15G}.
The adopted metallicity (i.e., the solar one) is not typical for usual Seyfert galaxies 
(whose NLR metallicity is generally higher than the solar metallicity), possibly nor BPT-valley AGNs 
(that could have sub-solar metallicity). 
However, as described below, it is useful to fix the metallicity to examine whether the ionization 
parameter alone can account for the difference in the emission-line flux ratios between 
BPT-valley objects and Seyferts. 
For the input SED, we adopt the following one:
\begin{eqnarray}
f_{\nu}={\nu}^{\rm \alpha_{UV}} \exp \left( -\frac{h\nu}{kT_{\rm BB}}\right) \exp \left( -\frac{kT_{\rm IR}}{h\nu}\right) 
+ a{\nu}^{\alpha_{\rm  X}}
\end{eqnarray}
as a typical spectrum of AGNs (see \citealt{1996hbic.book.....F}). 
$kT_{\rm IR}$ is the infrared cutoff of the big-blue bump, and we adopt 
$kT_{\rm IR}=0.01$ ryd \citep[see][]{1996hbic.book.....F}. 
$\alpha_{\rm UV}$ is the slope of the low-energy side of the big-blue bump.
We adopt $\alpha_{\rm UV} = 0.5$, which is typical for AGNs 
\citep{1996hbic.book.....F}. 
$\alpha_{\rm ox}$ is the UV--to--X-ray spectral slope, which determines the parameter $a$ in equation (6).
We adopt $\alpha_{\rm ox}=-1.35$, which is the average value of nearby Seyfert 1 galaxies
\citep[see][]{1993A&A...274..105W}.
$\alpha_{\rm x}$ is the X-ray slope, and we adopt $\alpha_{\rm x}=-0.85$ (see \citealt{2001ApJ...546..744N}).
$T_{\rm BB}$ is the characterizing the shape of the big-blue bump, and we adopt 490,000 K 
(see \citealt{2001ApJ...546..744N}).
The calculations end at the depth where the temperature falls to 3,000 K, 
below which gas does not contribute significantly to the flux of optical emission lines.

\begin{figure}[h]
 \centering
 \includegraphics[width=8.5cm]{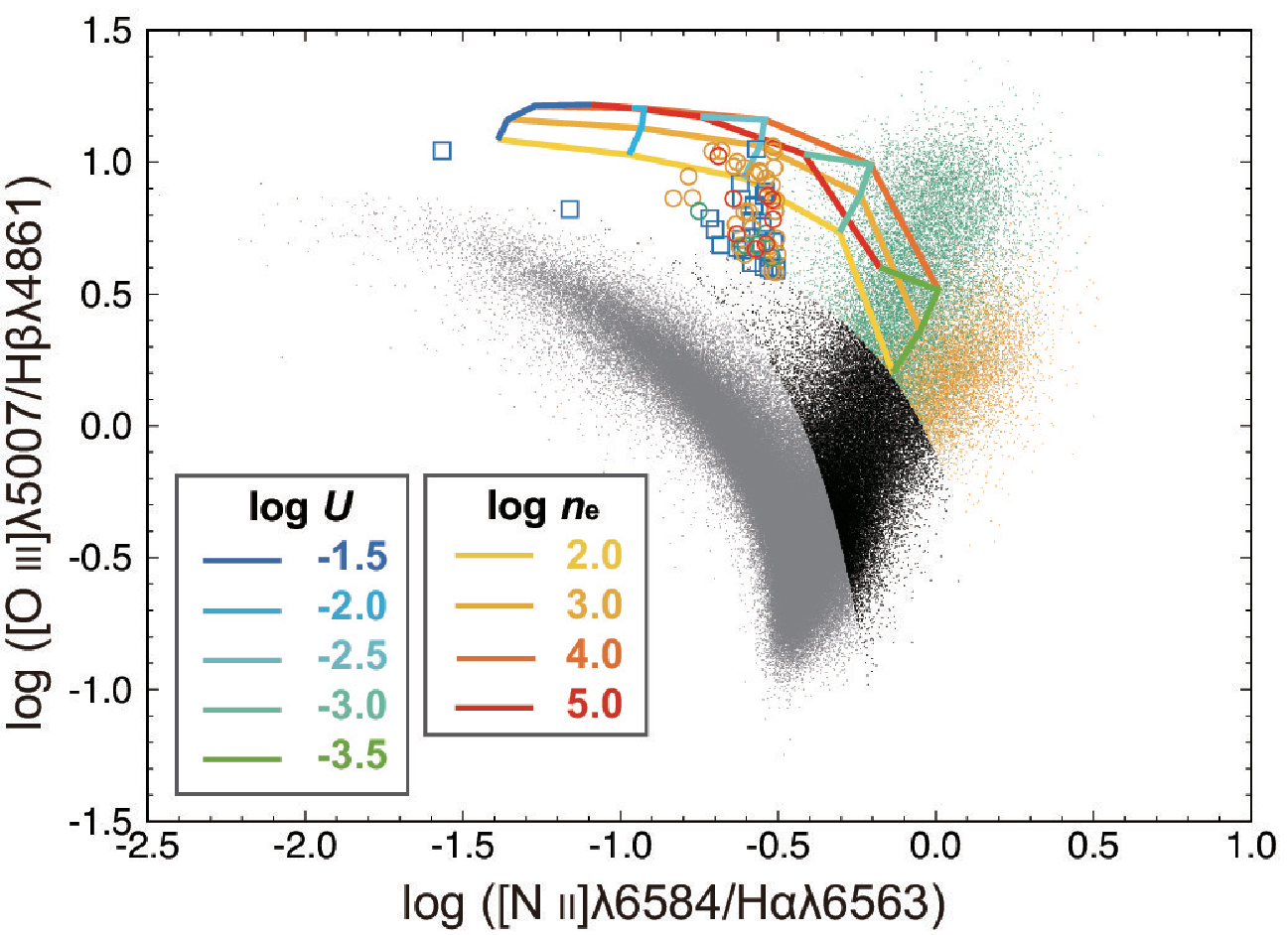}
 \caption{Same as Figure~\ref{BPT_classification} (without the inset panel in 
 Figure~\ref{BPT_classification}) but grids of photoionization models are overlaid. 
 Different colors of lines denote different parameters adopted in the calculations, 
 as shown in the inset panels.
 }
 \label{cloudy_1} 
\end{figure}

Figure~\ref{cloudy_1} shows the results of the photoionization model calculations, overlaid 
on the BPT diagram. 
Though the density effect is not significant in the range of 
$10^1$ cm$^{-3}$ $<$ $n_{\rm H}$ $<$ $10^5$ cm$^{-3}$, 
we can see the effect of the collisional de-excitation at $n_{\rm H}$ $>$ $10^4$ cm$^{-3}$. 
However, this figure suggests that the difference in the [N~{\sc ii}]$\lambda$6584/H$\alpha$$\lambda$6583 
flux ratio is more easily explained by the difference in the ionization parameter rather than by the difference 
in the gas density. 
More specifically, a higher ionization parameter by 0.5--1 dex in the BPT-valley objects with respect to 
the Seyfert sample is required to explain the lower [N~{\sc ii}]$\lambda$6584/H$\alpha$$\lambda$6583 
flux ratio of the BPT-valley objects. 

\begin{figure}[h]
\centering
 \includegraphics[width=8.5cm]{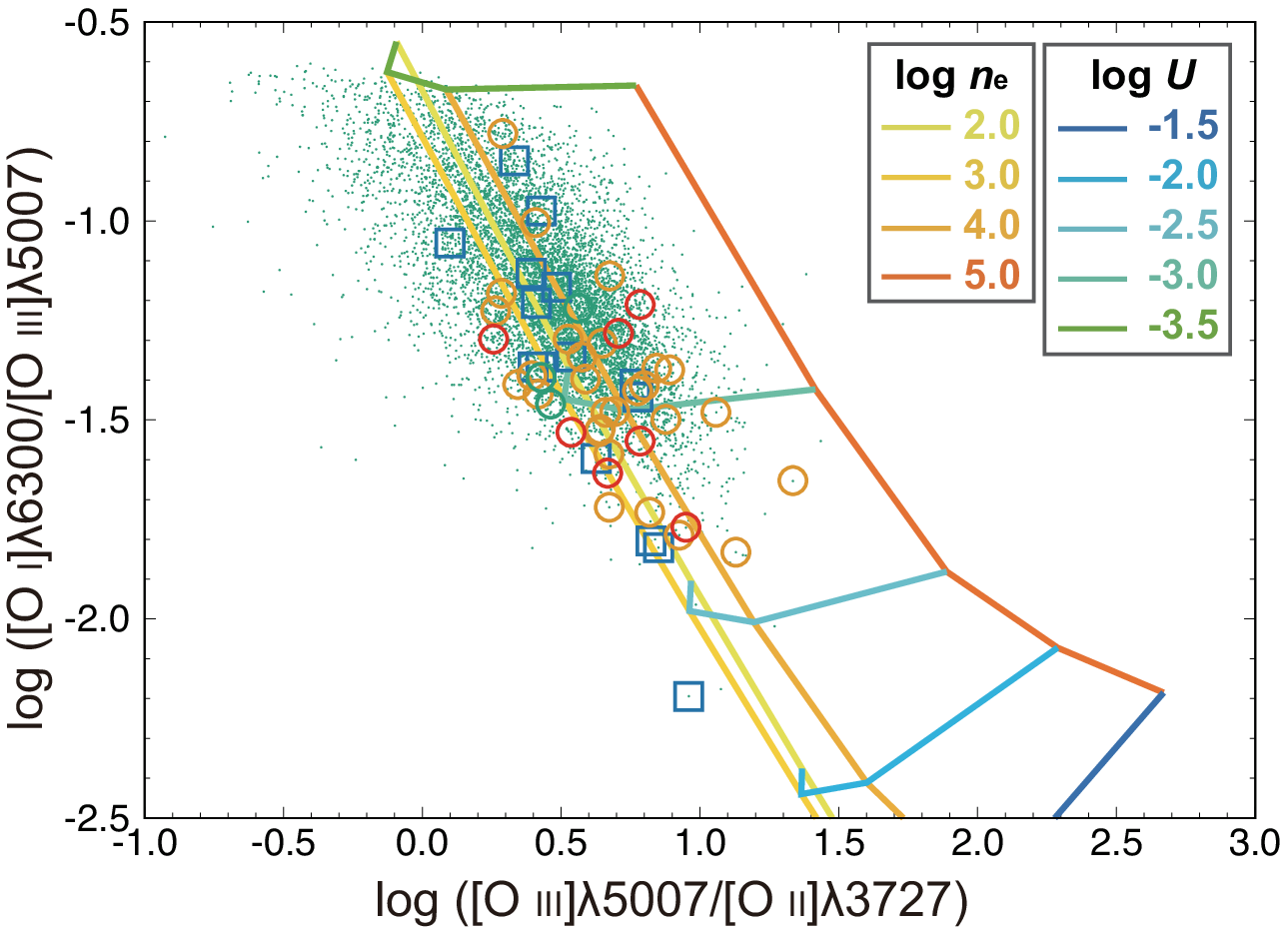}
 \caption{Diagnostic diagram of [O~{\sc iii}]$\lambda$5007/[O~{\sc ii}]$\lambda$3727 versus 
 [O~{\sc i}]$\lambda$6300/[O~{\sc iii}]$\lambda$5007. 
 The symbols and lines are the same as in Figure~\ref{cloudy_1}. 
 Note that only the BPT valley objects with S/N $>$ 5 of the [O~{\sc ii}]$\lambda$3727, 
 [O~{\sc iii}]$\lambda$5007 and [O~{\sc i}]$\lambda$6300 line are plotted.}
 \label{cloudy_2} 
\end{figure}

For examining whether the BPT-valley objects have a higher ionization parameter than the Seyfert sample, 
we investigate another diagnostic diagram that consists of the emission-line flux ratios of 
[O~{\sc iii}]$\lambda$5007/[O~{\sc ii}]$\lambda$3727 and [O~{\sc i}]$\lambda$6300/[O~{\sc iii}]$\lambda$5007 
(Figure~\ref{cloudy_2}). 
This diagram is useful to examine the effect of ionization parameter without suffering from 
the metallicity effect, because only oxygen lines are used and thus less sensitive to the metallicity. 
Figure~\ref{cloudy_2} shows that the BPT-valley sample and Seyfert sample have a similar gas density, 
that is consistent with our analysis presented in Section 4.1. 
More interestingly, Figure~\ref{cloudy_2} shows that the BPT-valley sample shows a systematically 
higher ionization parameter than the Seyfert sample, but the inferred difference in the ionization 
parameters is only less than 0.5 dex. 
This strongly suggests that the lower [N~{\sc ii}]$\lambda$6584/H$\alpha$$\lambda$6563 flux ratio 
in the BPT-valley sample with respect to the Seyfert sample is not explained by the ionization parameter 
(nor the gas density, as described in Section 4.1). 
Therefore we conclude that the BPT-valley AGNs are characterized by a systematlcally lower metallicity 
than the Seyfert sample, as originally proposed by \citet{2006MNRAS.371.1559G}.

\section{Disccusions}
As mentioned Section 1, low-metallicity AGNs are interesting to study the early phase of 
the AGN evolution.
However low-metallicity AGNs are very rare, so that little has been reported on
physical property of low-metallicity AGNs. 
In this section, we present some basic properties of BPT-valley objects which are expected 
to be low-metallicity AGNs.

\subsection{Stellar mass}
\begin{figure}[h]
\centering
 \includegraphics[width=8.5cm]{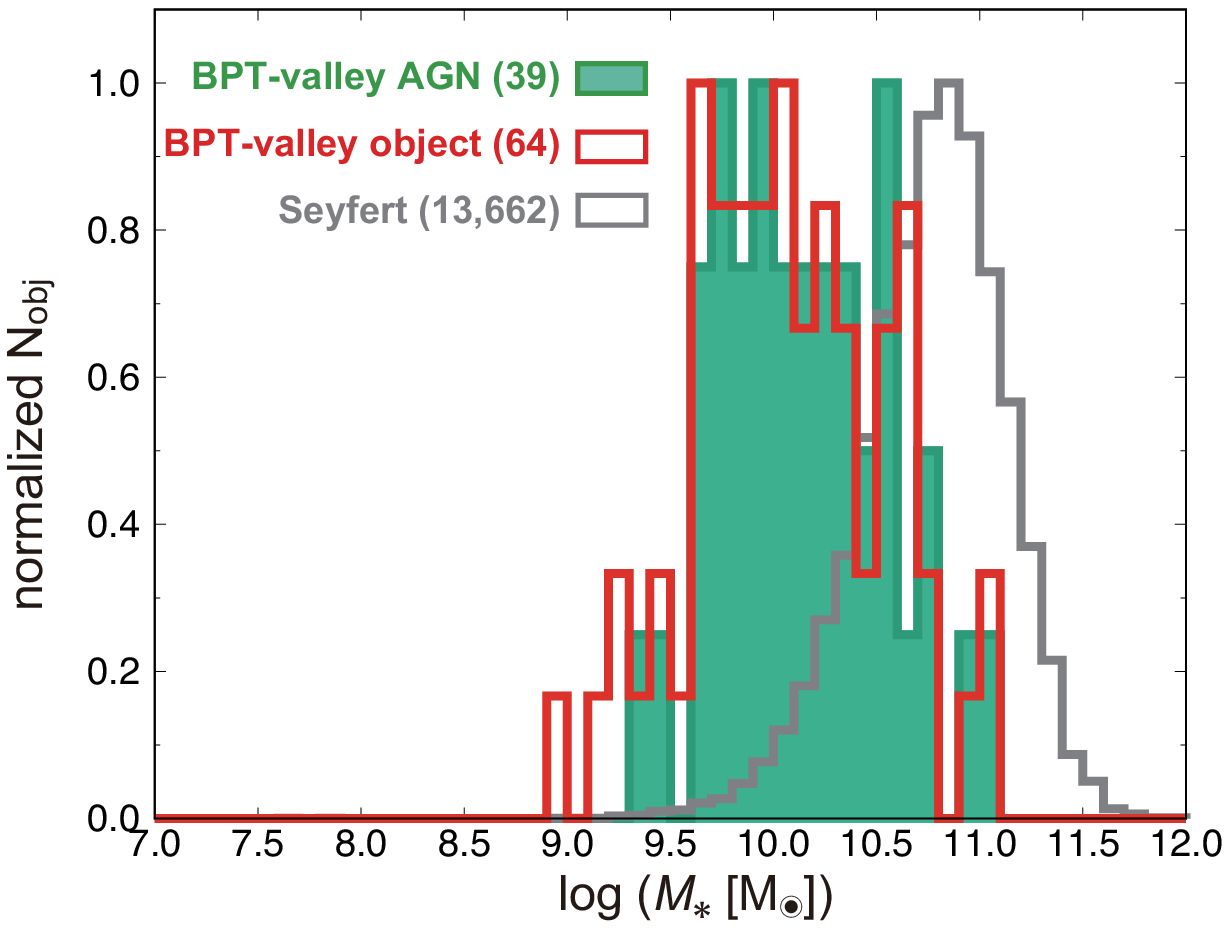}
 \caption{
 Same as Figure~\ref{SII_electron} but for the stellar mass.
 }
 \label{mass} 
\end{figure}

Naively it is expected that the stellar mass of low-metallicity AGNs is expected to be 
relatively low, as suggested by the mass-metallicity relation seen in star-forming galaxies 
\citep[e.g.,][]{2004ApJ...613..898T, 2006ApJ...647..970L}.
Accordingly \citet[][]{2006MNRAS.371.1559G} introduced a mass criterion 
(i.e., $M_{*} < 10^{10}\ {\rm M_{\odot}}$) to select low-metallicity AGNs. 
However, it is not clarified whether low-metallicity AGNs should be always found in 
a sample of AGNs with a low-mass host galaxy. 
Therefore, in this paper, we select low-metallicity AGNs without stellar-mass cut and 
investigate the mass distribution of host galaxies of low-metallicity AGNs. 
Here the stellar mass has been measured and given in the MPA-JHU DR7 catalog 
\cite[see also][]{2003MNRAS.341...33K}.
Among the 43 BPT-valley AGNs and 70 BPT-valley objects, the host mass is 
available for 39 and 64 objects, respectively. 
Figure~\ref{mass} shows the histogram of the stellar mass of the 39 BPT-valley AGNs, 
64 BPT-valley objects and 13,662 Seyferts.
The median of the stellar mass of the BPT-valley AGNs, BPT-valley objects 
and Seyferts are $10^{10.15}\ {\rm M_{\odot}}$, $10^{10.07}\ {\rm M_{\odot}}$ and 
$10^{10.77}\ {\rm M_{\odot}}$, respectively. 
This result clearly shows that the stellar mass of the BPT-valley AGNs is systematically lower 
than that of Seyferts. 
However, interestingly, a substantial fraction of the BPT-valley AGN (23 among 39 objects) 
are actually hosted by galaxies with $M_{*} > 10^{10}\ {\rm M_{\odot}}$, 
suggesting that low-metallicity AGNs are not necessarily hosted by low-mass galaxies. 
Note that such low-metallicity AGNs with a relatively massive host galaxy cannot be selected 
by the criteria of \citet[][]{2006MNRAS.371.1559G} due to the mass criterion of 
$M_{*} < 10^{10}\ {\rm M_{\odot}}$. 
Such low-metallicity AGNs hosted by a relatively massive host galaxy may be realized by 
taking into account of the inflow of low-metallicity gas from the surrounding environment  
\citep[e.g.,][]{2011A&A...535A..72H}.

\subsection{Electron temperature}
\begin{figure}[h]
\centering
 \includegraphics[width=8.5cm]{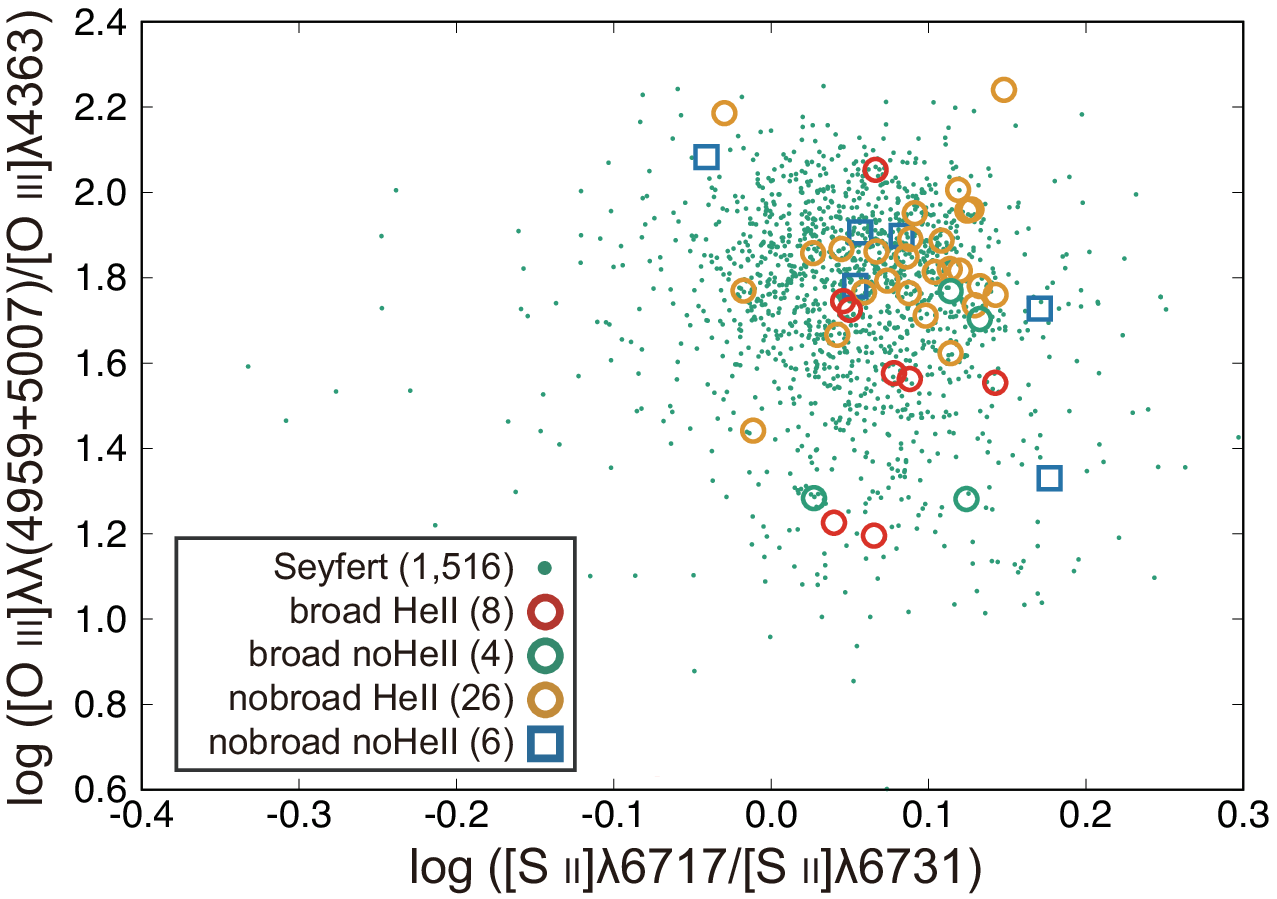}
 \caption{
 Diagnostic diagram of [O~{\sc iii}]$\lambda$$\lambda$(4959+5007)/
 [O~{\sc iii}]$\lambda$4363 versus [S~{\sc ii}]$\lambda$6717/[S~{\sc ii}]$\lambda$6731. 
 The symbols are the same as in Figure~\ref{BPT_classification}. 
 Note that only the BPT valley objects with S/N $>$ 3 of the [O~{\sc iii}]$\lambda$4363 line are plotted. 
 }
 \label{temperature_density} 
\end{figure}

Considering the effect of the metal cooling, 
low-metallicity AGNs are expected to be characterized by the higher electron temperature.
Hence we investigate the [O~{\sc iii}]$\lambda \lambda(4959+5007)$/[O~{\sc iii}]$\lambda4363$ line 
ratio which is very sensitive to the gas temperature.
Here it should be mentioned that, [O~{\sc iii}]$\lambda \lambda(4959+5007)$/[O~{\sc iii}]$\lambda4363$ line 
ratio also depends on the electron density \citep[see, e.g.,][]{2001ApJ...549..155N}. 
Therefore we investigate the  [O~{\sc iii}]$\lambda \lambda(4959+5007)$/[O~{\sc iii}]$\lambda4363$ and 
[S~{\sc ii}]$\lambda6717$/[S~{\sc ii}]$\lambda6731$ line ratios simultaneously in Figure~\ref{temperature_density}. 
Here this figure shows the emission-line flux ratios of BPT-valley objects and Seyferts but only for 
objects with a significant detection of the [O~{\sc iii}]$\lambda4363$ line (S/N $> 3$).
As described in Section 4.2, only objects with log ([O~{\sc iii}]$\lambda$5007/H$\beta$) $> 0.5$ are used 
(that results in 9,043 Seyferts and 70 BPT-valley objects).
Note that [O~{\sc iii}]$\lambda \lambda(4959+5007)$/[O~{\sc iii}]$\lambda4363$ line ratio is corrected 
for the reddening effect in the same way as Section 4.2.
The median values of log ([S~{\sc ii}]$\lambda6717$/[S~{\sc ii}]$\lambda6731$) of the BPT-valley AGNs, 
BPT-valley objects and Seyferts with a [O~{\sc iii}]$\lambda4363$ detection are 0.088, 0.088 and 0.055, 
respectively.
Therefore the electron density of the BPT-valley sample is slightly higher than that of Seyferts 
as already mentioned in Section 4.1. 
The median of log ([O~{\sc iii}]$\lambda \lambda(4959+5007)$/[O~{\sc iii}]$\lambda4363$) of 
the BPT-valley AGN, BPT-valley objects and Seyferts are 1.77, 1.77 and 1.79, respectively.
This result suggests that the electron temperature of the BPT-valley objects is not significantly higher 
than that of Seyferts. 
However, the fraction of objects showing a significant (S/N $> 3$) [O~{\sc iii}]$\lambda$4363 emission 
is very different between the Seyferts and BPT-valley objects. 
More specifically, 44 among the 70 BPT-valley objects ($\sim 63\ \%$) show the [O~{\sc ii}]$\lambda$4363 
emission while only 1,516 among 9,043 Seyferts ($\sim 17\ \%$) show the [O~{\sc iii}]$\lambda4363$ line. 
This difference infers that generally the gas temperature of the NLR in BPT-velley objects tends to be 
so high that the [O~{\sc iii}] $\lambda$4363 line is detected in most cases, 
while the typical gas temperature of the NLR in Seyferts may be lower than that in BPT-valley objects 
and only highly biased objects with a relatively high temperature in the Seyfert sample show the 
[O~{\sc iii}]$\lambda4363$ line. 
This result is consistent to our expectation that the BPT-valley objects is actually characterized by a 
relatively high gas temperature, due to the low gas metallicity.

\section{Conclusions}

In this paper, we focus on low-metallicity AGNs ($Z_{\rm NLR}$ $\lesssim$ $1\ Z_{\odot}$) 
which are very rare but important since they are in the early phase of the galaxy-SMBH co-evolution.
Specifically, in this work it is examined whether the BPT-valley selection is an effective and reliable 
way to identify low-metallicity AGNs, as proposed by \citet{2006MNRAS.371.1559G}. 
The main results are as follows:
\begin{itemize}
 \item We select 70 BPT valley sample which expected low metallicity AGN from 
14,253 Seyfert galaxies of MPA-JHU SDSS DR7 galaxy catalog.
 \item Out of 70 BPT-valley objects, 43 objects show clear evidence of the AGN based on 
 the detection of the broad H$\alpha$ component and/or He~{\sc ii}$\lambda$4686 emission.
  \item The typical gas density of the BPT-valley sample ($\sim$210 cm$^{-3}$) is not higher than that of 
  the Seyfert sample ($\sim$270 cm$^{-3}$), suggesting that the lower 
  [N~{\sc ii}]$\lambda$6584/H$\alpha$$\lambda$6563 ratio in the BPT-valley AGNs with respect to 
  the Seyfert sample is not caused by the collisional de-excitation effect. 
  \item The higher [O~{\sc iii}]$\lambda$5007/[O~{\sc ii}]$\lambda$3727 ratio in the BPT-valley sample 
  ($\sim$4.5) with respect to that in the Seyfert sample ($\sim$2.9) suggests a typically higher 
  ionization parameter of the BPT-valley sample; however, photoionization models suggest that 
  the inferred difference in the ionization parameter between the BPT-valley sample and 
  Seyfert sample is not enough to explain the observed lower 
  [N~{\sc ii}]$\lambda$6584/H$\alpha$$\lambda$6563 ratio of the BPT-valley sample.  
  \item The BPT-valley selection for identifying low-metallicity AGNs is thus confirmed to be a useful method; 
  in our analysis, more than 60\% of the BPT-valley sample are low-metallicity AGNs 
  ($Z_{\rm NLR}$ $\lesssim$ $1\ Z_{\odot}$). 

\end{itemize}

\acknowledgments

We would like to thank the anonymous referee for her/his
careful reading this paper and useful suggestions, and also
Masaru Kajisawa and Kazuyuki Ogura for their useful comments.
TN is financially supported by JSPS grants Nos. 25707010, 16H01101, and 16H03958. 
KM is also supported by JSPS grant No. 14J01811. 
Funding for the SDSS and SDSS-II has been provided by the Alfred P. Sloan Foundation, the Participating Institutions, the National Science Foundation, the U.S. Department of Energy, the National Aeronautics and Space Administration, the Japanese Monbukagakusho, the Max Planck Society, and the Higher Education Funding Council for England. The SDSS Web Site is http://www.sdss.org/.
The SDSS is managed by the Astrophysical Research Consortium for the Participating Institutions. The Participating Institutions are the American Museum of Natural History, Astrophysical Institute Potsdam, University of Basel, University of Cambridge, Case Western Reserve University, University of Chicago, Drexel University, Fermilab, the Institute for Advanced Study, the Japan Participation Group, Johns Hopkins University, the Joint Institute for Nuclear Astrophysics, the Kavli Institute for Particle Astrophysics and Cosmology, the Korean Scientist Group, the Chinese Academy of Sciences (LAMOST), Los Alamos National Laboratory, the Max-Planck-Institute for Astronomy (MPIA), the Max-Planck-Institute for Astrophysics (MPA), New Mexico State University, Ohio State University, University of Pittsburgh, University of Portsmouth, Princeton University, the United States Naval Observatory, and the University of Washington.

\bibliography{reference}

\begin{thebibliography}{}
\expandafter\ifx\csname natexlab\endcsname\relax\def\natexlab#1{#1}\fi

\bibitem[{{Abazajian} {et~al.}(2009){Abazajian}, {Adelman-McCarthy},
  {Ag{\"u}eros}, {Allam}, {Allende Prieto}, {An}, {Anderson}, {Anderson},
  {Annis}, {Bahcall}, \& et~al.}]{2009ApJS..182..543A}
{Abazajian}, K.~N., {Adelman-McCarthy}, J.~K., {Ag{\"u}eros}, M.~A., {et~al.}
  2009, \apjs, 182, 543

\bibitem[{{Baldwin} {et~al.}(1981){Baldwin}, {Phillips}, \&
  {Terlevich}}]{1981PASP...93....5B}
{Baldwin}, J.~A., {Phillips}, M.~M., \& {Terlevich}, R. 1981, \pasp, 93, 5

\bibitem[{{Cardelli} {et~al.}(1989){Cardelli}, {Clayton}, \&
  {Mathis}}]{1989ApJ...345..245C}
{Cardelli}, J.~A., {Clayton}, G.~C., \& {Mathis}, J.~S. 1989, \apj, 345, 245

\bibitem[{{Carniani} {et~al.}(2013){Carniani}, {Marconi}, {Biggs}, {Cresci},
  {Cupani}, {D'Odorico}, {Humphreys}, {Maiolino}, {Mannucci}, {Molaro},
  {Nagao}, {Testi}, \& {Zwaan}}]{2013A&A...559A..29C}
{Carniani}, S., {Marconi}, A., {Biggs}, A., {et~al.} 2013, \aap, 559, A29

\bibitem[{{Crenshaw} \& {Kraemer}(2000)}]{2000ApJ...532L.101C}
{Crenshaw}, D.~M., \& {Kraemer}, S.~B. 2000, \apjl, 532, L101

\bibitem[{{Ferland}(1996)}]{1996hbic.book.....F}
{Ferland}, G.~J. 1996, {Hazy, A Brief Introduction to Cloudy 90}

\bibitem[{{Ferland} {et~al.}(1998){Ferland}, {Korista}, {Verner}, {Ferguson},
  {Kingdon}, \& {Verner}}]{1998PASP..110..761F}
{Ferland}, G.~J., {Korista}, K.~T., {Verner}, D.~A., {et~al.} 1998, \pasp, 110,
  761

\bibitem[{{Grevesse} \& {Anders}(1989)}]{1989AIPC..183....1G}
{Grevesse}, N., \& {Anders}, E. 1989, in American Institute of Physics
  Conference Series, Vol. 183, Cosmic Abundances of Matter, ed. C.~J.
  {Waddington}, 1--8

\bibitem[{{Grevesse} \& {Noels}(1993)}]{1993oee..conf...15G}
{Grevesse}, N., \& {Noels}, A. 1993, in Origin and Evolution of the Elements,
  ed. N.~{Prantzos}, E.~{Vangioni-Flam}, \& M.~{Casse}, 15--25

\bibitem[{{Groves} {et~al.}(2006){Groves}, {Heckman}, \&
  {Kauffmann}}]{2006MNRAS.371.1559G}
{Groves}, B.~A., {Heckman}, T.~M., \& {Kauffmann}, G. 2006, \mnras, 371, 1559

\bibitem[{{Hayashi} {et~al.}(2015){Hayashi}, {Ly}, {Shimasaku}, {Motohara},
  {Malkan}, {Nagao}, {Kashikawa}, {Goto}, \& {Naito}}]{2015PASJ...67...80H}
{Hayashi}, M., {Ly}, C., {Shimasaku}, K., {et~al.} 2015, \pasj, 67, 80

\bibitem[{{Heckman}(1980)}]{1980A&A....87..152H}
{Heckman}, T.~M. 1980, \aap, 87, 152

\bibitem[{{Heckman} {et~al.}(1984){Heckman}, {Miley}, \&
  {Green}}]{1984ApJ...281..525H}
{Heckman}, T.~M., {Miley}, G.~K., \& {Green}, R.~F. 1984, \apj, 281, 525

\bibitem[{{Ho} {et~al.}(1997){Ho}, {Filippenko}, \&
  {Sargent}}]{1997ApJS..112..315H}
{Ho}, L.~C., {Filippenko}, A.~V., \& {Sargent}, W.~L.~W. 1997, \apjs, 112, 315

\bibitem[{{Husemann} {et~al.}(2011){Husemann}, {Wisotzki}, {Jahnke}, \&
  {S{\'a}nchez}}]{2011A&A...535A..72H}
{Husemann}, B., {Wisotzki}, L., {Jahnke}, K., \& {S{\'a}nchez}, S.~F. 2011,
  \aap, 535, A72

\bibitem[{{Izotov} \& {Thuan}(2008)}]{2008ApJ...687..133I}
{Izotov}, Y.~I., \& {Thuan}, T.~X. 2008, \apj, 687, 133

\bibitem[{{Kauffmann} {et~al.}(2003{\natexlab{a}}){Kauffmann}, {Heckman},
  {White}, {Charlot}, {Tremonti}, {Brinchmann}, {Bruzual}, {Peng}, {Seibert},
  {Bernardi}, {Blanton}, {Brinkmann}, {Castander}, {Cs{\'a}bai}, {Fukugita},
  {Ivezic}, {Munn}, {Nichol}, {Padmanabhan}, {Thakar}, {Weinberg}, \&
  {York}}]{2003MNRAS.341...33K}
{Kauffmann}, G., {Heckman}, T.~M., {White}, S.~D.~M., {et~al.}
  2003{\natexlab{a}}, \mnras, 341, 33

\bibitem[{{Kauffmann} {et~al.}(2003{\natexlab{b}}){Kauffmann}, {Heckman},
  {Tremonti}, {Brinchmann}, {Charlot}, {White}, {Ridgway}, {Brinkmann},
  {Fukugita}, {Hall}, {Ivezi{\'c}}, {Richards}, \&
  {Schneider}}]{2003MNRAS.346.1055K}
{Kauffmann}, G., {Heckman}, T.~M., {Tremonti}, C., {et~al.} 2003{\natexlab{b}},
  \mnras, 346, 1055

\bibitem[{{Kawakatu} {et~al.}(2003){Kawakatu}, {Umemura}, \&
  {Mori}}]{2003ApJ...583...85K}
{Kawakatu}, N., {Umemura}, M., \& {Mori}, M. 2003, \apj, 583, 85

\bibitem[{{Kewley} {et~al.}(2013){Kewley}, {Dopita}, {Leitherer}, {Dav{\'e}},
  {Yuan}, {Allen}, {Groves}, \& {Sutherland}}]{2013ApJ...774..100K}
{Kewley}, L.~J., {Dopita}, M.~A., {Leitherer}, C., {et~al.} 2013, \apj, 774,
  100

\bibitem[{{Kewley} {et~al.}(2001){Kewley}, {Dopita}, {Sutherland}, {Heisler},
  \& {Trevena}}]{2001ApJ...556..121K}
{Kewley}, L.~J., {Dopita}, M.~A., {Sutherland}, R.~S., {Heisler}, C.~A., \&
  {Trevena}, J. 2001, \apj, 556, 121

\bibitem[{{Kewley} {et~al.}(2006){Kewley}, {Groves}, {Kauffmann}, \&
  {Heckman}}]{2006MNRAS.372..961K}
{Kewley}, L.~J., {Groves}, B., {Kauffmann}, G., \& {Heckman}, T. 2006, \mnras,
  372, 961

\bibitem[{{Kojima} {et~al.}(2016){Kojima}, {Ouchi}, {Nakajima}, {Shibuya},
  {Harikane}, \& {Ono}}]{2016arXiv160503436K}
{Kojima}, T., {Ouchi}, M., {Nakajima}, K., {et~al.} 2016, ArXiv e-prints,
  arXiv:1605.03436

\bibitem[{{Komossa} \& {Schulz}(1997)}]{1997A&A...323...31K}
{Komossa}, S., \& {Schulz}, H. 1997, \aap, 323, 31

\bibitem[{{Kormendy} \& {Ho}(2013)}]{2013ARA&A..51..511K}
{Kormendy}, J., \& {Ho}, L.~C. 2013, \araa, 51, 511

\bibitem[{{Kriss}(1994)}]{1994ASPC...61..437K}
{Kriss}, G. 1994, in Astronomical Society of the Pacific Conference Series,
  Vol.~61, Astronomical Data Analysis Software and Systems III, ed. D.~R.
  {Crabtree}, R.~J. {Hanisch}, \& J.~{Barnes}, 437

\bibitem[{{Lamastra} {et~al.}(2010){Lamastra}, {Menci}, {Maiolino}, {Fiore}, \&
  {Merloni}}]{2010MNRAS.405...29L}
{Lamastra}, A., {Menci}, N., {Maiolino}, R., {Fiore}, F., \& {Merloni}, A.
  2010, \mnras, 405, 29

\bibitem[{{Lee} {et~al.}(2006){Lee}, {Skillman}, {Cannon}, {Jackson}, {Gehrz},
  {Polomski}, \& {Woodward}}]{2006ApJ...647..970L}
{Lee}, H., {Skillman}, E.~D., {Cannon}, J.~M., {et~al.} 2006, \apj, 647, 970

\bibitem[{{Magorrian} {et~al.}(1998){Magorrian}, {Tremaine}, {Richstone},
  {Bender}, {Bower}, {Dressler}, {Faber}, {Gebhardt}, {Green}, {Grillmair},
  {Kormendy}, \& {Lauer}}]{1998AJ....115.2285M}
{Magorrian}, J., {Tremaine}, S., {Richstone}, D., {et~al.} 1998, \aj, 115, 2285

\bibitem[{{Marconi} \& {Hunt}(2003)}]{2003ApJ...589L..21M}
{Marconi}, A., \& {Hunt}, L.~K. 2003, \apjl, 589, L21

\bibitem[{{Masters} {et~al.}(2014){Masters}, {McCarthy}, {Siana}, {Malkan},
  {Mobasher}, {Atek}, {Henry}, {Martin}, {Rafelski}, {Hathi}, {Scarlata},
  {Ross}, {Bunker}, {Blanc}, {Bedregal}, {Dom{\'{\i}}nguez}, {Colbert},
  {Teplitz}, \& {Dressler}}]{2014ApJ...785..153M}
{Masters}, D., {McCarthy}, P., {Siana}, B., {et~al.} 2014, \apj, 785, 153

\bibitem[{{Matsuoka} {et~al.}(2009){Matsuoka}, {Nagao}, {Maiolino}, {Marconi},
  \& {Taniguchi}}]{2009A&A...503..721M}
{Matsuoka}, K., {Nagao}, T., {Maiolino}, R., {Marconi}, A., \& {Taniguchi}, Y.
  2009, \aap, 503, 721

\bibitem[{{Mendoza}(1983)}]{1983IAUS..103..143M}
{Mendoza}, C. 1983, in IAU Symposium, Vol. 103, Planetary Nebulae, ed. D.~R.
  {Flower}, 143--172

\bibitem[{{Murayama} \& {Taniguchi}(1998)}]{1998ApJ...497L...9M}
{Murayama}, T., \& {Taniguchi}, Y. 1998, \apjl, 497, L9

\bibitem[{{Nagao} {et~al.}(2006{\natexlab{a}}){Nagao}, {Maiolino}, \&
  {Marconi}}]{2006A&A...459...85N}
{Nagao}, T., {Maiolino}, R., \& {Marconi}, A. 2006{\natexlab{a}}, \aap, 459, 85

\bibitem[{{Nagao} {et~al.}(2006{\natexlab{b}}){Nagao}, {Marconi}, \&
  {Maiolino}}]{2006A&A...447..157N}
{Nagao}, T., {Marconi}, A., \& {Maiolino}, R. 2006{\natexlab{b}}, \aap, 447,
  157

\bibitem[{{Nagao} {et~al.}(2002){Nagao}, {Murayama}, {Shioya}, \&
  {Taniguchi}}]{2002ApJ...567...73N}
{Nagao}, T., {Murayama}, T., {Shioya}, Y., \& {Taniguchi}, Y. 2002, \apj, 567,
  73

\bibitem[{{Nagao} {et~al.}(2001{\natexlab{a}}){Nagao}, {Murayama}, \&
  {Taniguchi}}]{2001ApJ...546..744N}
{Nagao}, T., {Murayama}, T., \& {Taniguchi}, Y. 2001{\natexlab{a}}, \apj, 546,
  744

\bibitem[{{Nagao} {et~al.}(2001{\natexlab{b}}){Nagao}, {Murayama}, \&
  {Taniguchi}}]{2001ApJ...549..155N}
---. 2001{\natexlab{b}}, \apj, 549, 155

\bibitem[{{Nagao} {et~al.}(2000){Nagao}, {Taniguchi}, \&
  {Murayama}}]{2000AJ....119.2605N}
{Nagao}, T., {Taniguchi}, Y., \& {Murayama}, T. 2000, \aj, 119, 2605

\bibitem[{{Nakajima} \& {Ouchi}(2014)}]{2014MNRAS.442..900N}
{Nakajima}, K., \& {Ouchi}, M. 2014, \mnras, 442, 900

\bibitem[{{Newman} {et~al.}(2014){Newman}, {Buschkamp}, {Genzel}, {F{\"o}rster
  Schreiber}, {Kurk}, {Sternberg}, {Gnat}, {Rosario}, {Mancini}, {Lilly},
  {Renzini}, {Burkert}, {Carollo}, {Cresci}, {Davies}, {Eisenhauer}, {Genel},
  {Shapiro Griffin}, {Hicks}, {Lutz}, {Naab}, {Peng}, {Tacconi}, {Wuyts},
  {Zamorani}, {Vergani}, \& {Weiner}}]{2014ApJ...781...21N}
{Newman}, S.~F., {Buschkamp}, P., {Genzel}, R., {et~al.} 2014, \apj, 781, 21

\bibitem[{{Osterbrock}(1989)}]{1989agna.book.....O}
{Osterbrock}, D.~E. 1989, {Astrophysics of gaseous nebulae and active galactic
  nuclei (Mill Valley: University Science Books)}

\bibitem[{{Reines} \& {Volonteri}(2015)}]{2015ApJ...813...82R}
{Reines}, A.~E., \& {Volonteri}, M. 2015, \apj, 813, 82

\bibitem[{{Schramm} {et~al.}(2008){Schramm}, {Wisotzki}, \&
  {Jahnke}}]{2008A&A...478..311S}
{Schramm}, M., {Wisotzki}, L., \& {Jahnke}, K. 2008, \aap, 478, 311

\bibitem[{{Schulze} \& {Wisotzki}(2011)}]{2011A&A...535A..87S}
{Schulze}, A., \& {Wisotzki}, L. 2011, \aap, 535, A87

\bibitem[{{Shapley} {et~al.}(2015){Shapley}, {Reddy}, {Kriek}, {Freeman},
  {Sanders}, {Siana}, {Coil}, {Mobasher}, {Shivaei}, {Price}, \& {de
  Groot}}]{2015ApJ...801...88S}
{Shapley}, A.~E., {Reddy}, N.~A., {Kriek}, M., {et~al.} 2015, \apj, 801, 88

\bibitem[{{Shirazi} {et~al.}(2014){Shirazi}, {Brinchmann}, \&
  {Rahmati}}]{2014ApJ...787..120S}
{Shirazi}, M., {Brinchmann}, J., \& {Rahmati}, A. 2014, \apj, 787, 120

\bibitem[{{Steidel} {et~al.}(2014){Steidel}, {Rudie}, {Strom}, {Pettini},
  {Reddy}, {Shapley}, {Trainor}, {Erb}, {Turner}, {Konidaris}, {Kulas}, {Mace},
  {Matthews}, \& {McLean}}]{2014ApJ...795..165S}
{Steidel}, C.~C., {Rudie}, G.~C., {Strom}, A.~L., {et~al.} 2014, \apj, 795, 165

\bibitem[{{Tremonti} {et~al.}(2004){Tremonti}, {Heckman}, {Kauffmann},
  {Brinchmann}, {Charlot}, {White}, {Seibert}, {Peng}, {Schlegel}, {Uomoto},
  {Fukugita}, \& {Brinkmann}}]{2004ApJ...613..898T}
{Tremonti}, C.~A., {Heckman}, T.~M., {Kauffmann}, G., {et~al.} 2004, \apj, 613,
  898

\bibitem[{{van Zee} {et~al.}(1998){van Zee}, {Salzer}, \&
  {Haynes}}]{1998ApJ...497L...1V}
{van Zee}, L., {Salzer}, J.~J., \& {Haynes}, M.~P. 1998, \apjl, 497, L1

\bibitem[{{Vanden Berk} {et~al.}(2001){Vanden Berk}, {Richards}, {Bauer},
  {Strauss}, {Schneider}, {Heckman}, {York}, {Hall}, {Fan}, {Knapp},
  {Anderson}, {Annis}, {Bahcall}, {Bernardi}, {Briggs}, {Brinkmann}, {Brunner},
  {Burles}, {Carey}, {Castander}, {Connolly}, {Crocker}, {Csabai}, {Doi},
  {Finkbeiner}, {Friedman}, {Frieman}, {Fukugita}, {Gunn}, {Hennessy},
  {Ivezi{\'c}}, {Kent}, {Kunszt}, {Lamb}, {Leger}, {Long}, {Loveday}, {Lupton},
  {Meiksin}, {Merelli}, {Munn}, {Newberg}, {Newcomb}, {Nichol}, {Owen}, {Pier},
  {Pope}, {Rockosi}, {Schlegel}, {Siegmund}, {Smee}, {Snir}, {Stoughton},
  {Stubbs}, {SubbaRao}, {Szalay}, {Szokoly}, {Tremonti}, {Uomoto}, {Waddell},
  {Yanny}, \& {Zheng}}]{2001AJ....122..549V}
{Vanden Berk}, D.~E., {Richards}, G.~T., {Bauer}, A., {et~al.} 2001, \aj, 122,
  549

\bibitem[{{Walter} \& {Fink}(1993)}]{1993A&A...274..105W}
{Walter}, R., \& {Fink}, H.~H. 1993, \aap, 274, 105

\bibitem[{{Wang} {et~al.}(2010){Wang}, {Carilli}, {Neri}, {Riechers}, {Wagg},
  {Walter}, {Bertoldi}, {Menten}, {Omont}, {Cox}, \&
  {Fan}}]{2010ApJ...714..699W}
{Wang}, R., {Carilli}, C.~L., {Neri}, R., {et~al.} 2010, \apj, 714, 699

\bibitem[{{Yabe} {et~al.}(2015){Yabe}, {Ohta}, {Akiyama}, {Bunker}, {Dalton},
  {Ellis}, {Glazebrook}, {Goto}, {Imanishi}, {Iwamuro}, {Okada}, {Shimizu},
  {Takato}, {Tamura}, {Tonegawa}, \& {Totani}}]{2015PASJ...67..102Y}
{Yabe}, K., {Ohta}, K., {Akiyama}, M., {et~al.} 2015, \pasj, 67, 102

\bibitem[{{York} {et~al.}(2000){York}, {Adelman}, {Anderson}, {Anderson},
  {Annis}, {Bahcall}, {Bakken}, {Barkhouser}, {Bastian}, {Berman}, {Boroski},
  {Bracker}, {Briegel}, {Briggs}, {Brinkmann}, {Brunner}, {Burles}, {Carey},
  {Carr}, {Castander}, {Chen}, {Colestock}, {Connolly}, {Crocker}, {Csabai},
  {Czarapata}, {Davis}, {Doi}, {Dombeck}, {Eisenstein}, {Ellman}, {Elms},
  {Evans}, {Fan}, {Federwitz}, {Fiscelli}, {Friedman}, {Frieman}, {Fukugita},
  {Gillespie}, {Gunn}, {Gurbani}, {de Haas}, {Haldeman}, {Harris}, {Hayes},
  {Heckman}, {Hennessy}, {Hindsley}, {Holm}, {Holmgren}, {Huang}, {Hull},
  {Husby}, {Ichikawa}, {Ichikawa}, {Ivezi{\'c}}, {Kent}, {Kim}, {Kinney},
  {Klaene}, {Kleinman}, {Kleinman}, {Knapp}, {Korienek}, {Kron}, {Kunszt},
  {Lamb}, {Lee}, {Leger}, {Limmongkol}, {Lindenmeyer}, {Long}, {Loomis},
  {Loveday}, {Lucinio}, {Lupton}, {MacKinnon}, {Mannery}, {Mantsch}, {Margon},
  {McGehee}, {McKay}, {Meiksin}, {Merelli}, {Monet}, {Munn}, {Narayanan},
  {Nash}, {Neilsen}, {Neswold}, {Newberg}, {Nichol}, {Nicinski}, {Nonino},
  {Okada}, {Okamura}, {Ostriker}, {Owen}, {Pauls}, {Peoples}, {Peterson},
  {Petravick}, {Pier}, {Pope}, {Pordes}, {Prosapio}, {Rechenmacher}, {Quinn},
  {Richards}, {Richmond}, {Rivetta}, {Rockosi}, {Ruthmansdorfer}, {Sandford},
  {Schlegel}, {Schneider}, {Sekiguchi}, {Sergey}, {Shimasaku}, {Siegmund},
  {Smee}, {Smith}, {Snedden}, {Stone}, {Stoughton}, {Strauss}, {Stubbs},
  {SubbaRao}, {Szalay}, {Szapudi}, {Szokoly}, {Thakar}, {Tremonti}, {Tucker},
  {Uomoto}, {Vanden Berk}, {Vogeley}, {Waddell}, {Wang}, {Watanabe},
  {Weinberg}, {Yanny}, {Yasuda}, \& {SDSS Collaboration}}]{2000AJ....120.1579Y}
{York}, D.~G., {Adelman}, J., {Anderson}, Jr., J.~E., {et~al.} 2000, \aj, 120,
  1579

\end{thebibliography}

\end{document}